\documentclass[12pt]{article}
\usepackage{amsmath,amssymb,mathtools}
\usepackage{graphicx,enumitem,color,threeparttable,booktabs,multirow,tikz,makecell}
\usetikzlibrary{arrows.meta,positioning}
\usepackage[title]{appendix}
\usepackage[hyphens]{url}
\usepackage{hyperref}
\newtheorem{theorem}{Theorem}

\allowdisplaybreaks[4]

\newcommand{\E}{\mathrm{E}}

\newcommand{\pto}{\stackrel{p}{\longrightarrow}}
\newcommand{\dto}{\stackrel{d}{\longrightarrow}}

\usepackage[round,comma]{natbib}
\newcommand{\anon}{1}

\begin{document}

\def\spacingset#1{\renewcommand{\baselinestretch}%
{#1}\small\normalsize} \spacingset{1}


\if1\anon
{
  \title{\bf Causal Inference in Panel Data with a Continuous Treatment}
  \author{Zhiguo Xiao\thanks{Corresponding Address: Room 709 Siyuan Faculty Building, 670 Guoshun Rd., Shanghai 200433, China. Email: zhiguo\_xiao@fudan.edu.cn. The authors gratefully acknowledge the financial support by the National Natural Science Foundation of China (Grant No. 72232002).}\hspace{.2cm}\\
    Department of Statistics and Data Science, Fudan University, China\\
    and \\
    Peikai Wu \\
    Department of Statistics and Data Science, Fudan University, China}
  \maketitle
} \fi

\if0\anon
{
  \bigskip
  \bigskip
  \bigskip
  \begin{center}
    {\LARGE\bf Causal Inference in Panel Data with a Continuous Treatment}
\end{center}
  \medskip
} \fi

\bigskip
\begin{abstract}
This paper proposes a framework that incorporates the two-way fixed effects model as a special case to conduct causal inference with a continuous treatment. Treatments are allowed to change over time and potential outcomes are dependent on historical treatments. Regression models on potential outcomes, along with the sequentially conditional independence assumptions (SCIAs) are introduced to identify the treatment effects, which are measured by aggre causal responses. Least squares and generalized method of moments (GMM) estimators are developed for model parameters, which are then used to estimate the aggregate causal effects. We establish the asymptotic properties of these aggregate estimators. Additionally, we propose employing directed acyclic graphs (DAGs) to test the validity of the SCIAs. An application examining the aid-growth relationship illustrates the proposed methodology.
\end{abstract}

\noindent%
\vfill

\newpage
\spacingset{1.8} 

\section{Introduction}
Causal inference with panel data, especially using the difference-in-differences (DiD) approach, has been given increasing attention recently. As \citet{de2023two} put it, the recent literature on DiD  ``has developed in such a quick and dynamic manner that some practitioners may have gotten lost in the whirlwind of new working papers''. This is not a surprise, as the default DiD approach, the two-way fixed effects (TWFE) method, has recently been shown to be unable to deliver correct causal effect estimates under general scenarios.

In the cross-sectional setup, \citet{angrist2009mostly}, citing \citet{yitzhaki1996using}, have warned of the limitation of the least squares (LS) in estimating causal effects. Namely, that the LS estimator, which is a weighted average of covariate-specific treatment effects, is in general not equal to the typically used causal estimand average treatment effect on the treated (ATET), which is also a weighted average of covariate-specific treatment effects, as they are using different weights. More importantly, ATET ``puts the most weight on
covariate cells containing those who are most likely to be treated. In contrast, regression puts the most weight on covariate cells where the conditional variance of treatment status is largest.'' Only when there is treatment effects homogeneity will the LS estimator equal ATET, as the difference in weighting then does not matter.

An array of papers, including \citet{de2020two}, \citet{goodman2021difference}, \citet{callaway2021difference}, \citet{sun2021estimating}, etc., have discovered similar issue again in panel data. More importantly, in panel data with multiple periods, some of the weights used by the TWFE estimator to combine cross-group treatment effects are negative, making it possible that when there is strong heterogeneity of treatment effects, the TWFE estimator could be negative whilst each of the cross-group treatment effects are positive, and vice versa. The literature has proposed a number of alternative causal estimands and estimators in the hope to rescue the beleaguered DiD/TWFE.

Most of the above research fix on a canonical DiD design, i.e., a binary treatment with staggered adoption, where treatment effects heterogeneity is mainly caused by difference in individuals' initial treatment times.

Another source of treatment heterogeneity is the multiplicity of treatment levels, which appears in a lot of forms in the real world, such as foreign aid, state level minimum wages, local public spendings, local taxes, subsidies, trade tariffs, firm R\&D investments, individual smoking intensity, precipitation, pollution, etc., are continuously measured. However, research articles focusing on dealing with continous treatment, and in particular with panel data, are relatively scarce, with \citet{callaway2024difference} and \citet{klosin2022estimating} being a few exceptions. \citet{callaway2024difference} extend the above discussion to continuous treatment with staggered adoption, but their framework allows only treatment to vary cross-sectionally. \citet{klosin2022estimating} allow treatment to change both across individuals and over time, but restricts to contemporaneous treatment effects and one-way fixed effects. A more realistic situation, and with even higher degree of treatment effects heterogeneity, is when treatment is continuous, can change both across individuals and over time, and having potential lagging effects. This is where this paper kicks in.

We build on \citet{sobel2012does}'s characterization of causal inference with panel data to propose a general framework for evaluating treatment effects with a continuous treatment. Individuals enter the study without previous treatment, and the treatment effects are non-anticipatory. Staring form the first period, individuals get treated by a continuous treament, and for any individual, treatment levels are allowed to be different at any two consecutive periods. At any time period $t$, potential outcomes are indexed by treatment history until $t$. That is to say, past treatment history, instead of current treatment alone, determines the potential outcomes.

We use average causal responses (ACRs) to an incremental change in the treatment, also known as the average derivatives, as the \textit{building blocks} of our aggregate treatment effects targets \citep[see][for related discussions]{angrist2000interpretation,rothenhausler2019incremental,callaway2024difference,chernozhukov2022automatic,klosin2022estimating}. Two types of structural causal models (i.e., models of the potential outcomes) and their accompanying identifying assumptions are introduced to produce two empirical models (i.e., models of observable outcomes) from which our aggregate treatment effects are to be obtained. Different from in the traditional TWFE models, the treatment term appears in our models in very general forms, thus enabling flexible characterization of treatment effects heterogeneity, with the simplest case reduces to the classical TWFE model. The aggregate treatment effects in our models are functions of two parts: parameters that can be estimated from our empirical models by generalized method of moments (GMM) for the model with weaker assumptions, and by LSDV/TWFE for the model with stronger assumptions, and moments of treatments that can be estimated by their sample analogs. The resulting aggregate treatment effects estimators are asymptotically normal and we suggest using nonparametric cross-sectional bootstrap to obtain their standard errors. Theoretical properties of the estimators are derived both for static and dynamic panel data models. 

Our identifying assumptions, namely, the sequentially conditional independence assumptions (SCIAs, also called the sequentially conditional ignorability assumptions) are the dynamic versions of the conditional independence assumption (CIA) used in the cross-sectional data. We test the validity of the SCIAs using directed acyclic graphs (DAGs), which are advocated by \citet{pearl2009causality}. As DAGs are formal representations of researchers' contextual knowledge, we believe that such a practice might help the research community to engage in more targeted debates.

We make two contributions to the on-going research on causal inference with panel data.

First, by entending \citet{sobel2012does}'s work and connecting it with the current research on the DiD/TWFE, we develop a general framework to conduct causal inference for panel data (both static and dynamic) with a continuous treatment that subsumes the two-way fixed effects as a special case. The framework has several distinctive features compared with the popular DiD/TWFE with staggered adoption. The treatment can move up or down multiple times. This is a feature that \citet{de2024difference} call for special attention.  Treatment effects are modeled through a flexible function of past treatment history, instead of just current period treatment. This feature provides modelers with more latitude in characterizing treatment effect heterogeneity. The estimators of aggregate treatment effects are no longer weighted averages of pairwise treatment effects.

Second, we introduce a new approach, namely, the DAGs, to test the validity of the SCIAs that are crucial in identifying the causal estimands, and we connect and compare this new approach with the traditional practice of imposing the parallel trend assumptions (PTAs) and testing with pre-treatment trend patterns. The SCIAs are essentially similar to the parallel trend assumptions. However, the DAG tests are different from the pre-treatment tests, as they do not rest on existing data. Rather, they translate the researcher's contextual knowledge into a structured graph through which conditional independence relationship can be uncovered by mathematical deduction. Since both SCIAs and PTAs, or any identifying assumptions that translate models on potential outcomes into models on observed outcomes, are assumption on the (conditional) distribution of potential outcomes which are only partially observable, the validity of any data-based test of them has to rely on additional knowledge and/or additional implicit assumptions, which themselves require verification as well. The DAG-based test solves the outcome data partiality problem by starting directly with the additional knowledge that is required for causal effects identification.  

The rest of the paper is organized as follows. Section 2 introduces the linear static panel data framework for causal inference with a continuous treatment. Section 3 specifies the functional forms of treatment effects heterogeneity and develops estimators for the aggregate treatment effects. Section 4 presents the theoretical properties of the causal estimators. Section 5 illustrates the application of the proposed method with a real example. Section 6 explains the differences between the proposed method and the canonical DiD/TWFE methods, and Section 7 concludes. Specific examples of functional form treatment effects heterogeneity, proofs of theoretical results, and results for dynamic panel models are contained in the Appendix.

\section{Model setup}

\subsection{Basic framework with a continuous treatment}
We follow \citet{sobel2012does} to consider a balanced panel data with $i=1,\dots, N$ and $t=1,\dots, T$ representing individual units and time periods, respectively. $Y_{it}, D_{it}, \mathbf{X}_{it}$ denote the observed outcome, the treatment and the vector of covariates for individual $i$ at time $t$, respectively.\footnote{The method in this paper works as well for unbalanced data. We choose the balanced panel setup to reduce the unnecessary burden of notation generality.} $U_{i}$ and $V_{t}$ represent unobservable individual fixed effects and time fixed effects respectively. We assume that $(Y_{i}, D_{i}, \mathbf{X}_{i}), ~i=1,\dots, N$ are i.i.d., where $Y_{i}=(Y_{i1}, \dots, Y_{iT})$, $D_{i}=(D_{i1}, \dots, D_{iT})$, and $\mathbf{X}_{i}=(\mathbf{X}_{i1}, \dots, \mathbf{X}_{iT})$.

Let $\mathcal{D}$ be the support of $D$, i.e., $\mathcal{D}$ is the collection of all possible values of $D$. In the literature, $\mathcal{D}$ is often set to be $\{0,1\}$. However, as pointed out by \citet{heckman2007econometric}, ``A two treatment environment receives the most attention in the theoretical literature, but the multiple treatment environment is the one most frequently encountered in practice.'' \citet{callaway2024difference} consider a scenario of $\mathcal{D}=\{0\} \cup [d_{L}, d_{U}]$ with $P(D=0)>0$ and $0<d_{L}<d_{U}<\infty$, which is suited for a staggered adoption design with positive fraction of never-treated units. Here we are interested in the situation where $D$ is a treatment intensity that varies across individuals and time with support set $\mathcal{D}=[0,d^{\dagger}]$, where $0$ denotes no treatment and $d^{\dagger}$ is a fixed positive number indicating the maximum possible treatment. More specifically, we assume that for any individual $i$ at any time period $t$, $D_{it}$ can take any values in $\mathcal{D}=[0,d^{\dagger}]$. Let $f_{t}(\cdot)$ be the probability density function of $D_{it}$, and we assume that $f_{t}(d)>0, \forall d\in \mathcal{D}=[0,d^{\dagger}]$. Our characterization of $\mathcal{D}$ is consistent with many continuous treatments, such as those mentioned in the beginning of the paper. We assume that no individual enters the study with prior treatment history, i.e.,
\[
D_{it}=0, \forall i=1,\dots, n;t=0,-1,-2,\dots
\]

Let $\mathcal{D}^{t}=\mathcal{D}\times \cdots \times\mathcal{D}$ be the support set of $(D_{i1}, \dots, D_{it})$.  For any $t=1,\dots, T$ and any variable $Z$, we define $\overline{Z}_{it}=(Z_{i1},\dots, Z_{it})$ and $\overline{Z}_{i\{j>t\}}=(Z_{i,t+1},\dots, Z_{iT})$. In particular, lets use $d$ to denote the value of $D$, and we have $\overline{d}_{it}=(d_{i1},\dots, d_{it})$, $\overline{d}_{i\{j>t\}}=(d_{i,t+1},\dots, d_{iT})$, and that
\begin{equation*}
  \overline{d}_{iT}=(\overline{d}_{it}, \overline{d}_{i\{j>t\}}), ~~\forall t=1, \dots, T-1.
\end{equation*}

We follow \citet[\S 19.2]{hernan2024causal} to call a complete treatment sequence $\overline{d}_{iT}\in \mathcal{D}^{T}$ a \textit{treatment strategy}. Under the treatment strategy $\overline{d}_{iT}$, the potential outcome of individual $i$ at time $t$ is denoted by $Y_{it}(\overline{d}_{iT})$, and the complete sequence of potential outcomes is $\overline{Y}_{i}(\overline{d}_{iT})=(Y_{i1}(\overline{d}_{iT}), \dots, Y_{iT}(\overline{d}_{iT}))$.

We assume that potential outcomes are not affected by future treatments. This assumption is known as the no-anticipation assumption. Specifically, given two treatment strategies $\overline{d}_{iT}=(\overline{d}_{it}, \overline{d}_{i\{j>t\}}), \overline{d}_{iT}^{*}=(\overline{d}_{it}^{*}, \overline{d}_{i\{j>t\}}^{*})$ such that $\overline{d}_{it}=\overline{d}_{it}^{*}$, we have that $Y_{it}(\overline{d}_{iT})=Y_{it}(\overline{d}_{iT}^{*})$. With the no-anticipation assumption, the potential outcome of individual $i$ at time $t$ under treatment strategy $\overline{d}_{iT}$ can be written as $Y_{it}(\overline{d}_{it})$.

For simplicity we also assume that $\mathbf{X}_{it}$ is not affected by past treatment history. This assumption is implicitly adopted by the econometric literature. Nevertheless, the main results of this paper can be easily generalized to allow for the situation where $\mathbf{X}_{it}$ is a function of $\overline{d}_{it}$.

\subsection{Causal estimands}
Now let's define the causal estimands. The treatment difference of two treatment history $\overline{d}_{it},\overline{d}_{it}^{*}$ on individual $i$ at time $t$ is
\begin{equation*}
  Y_{it}(\overline{d}_{it})-Y_{it}(\overline{d}_{it}^{*}),~~t=1,\dots, T,
\end{equation*}
which is unidentifiable unless we have parallel worlds. What we can identify and estimate in the real world is some kind of averages of the above individual level differences. A general form of such average is
\begin{equation}
  \label{estm1}
  ATE_{t}(\overline{d}_{it},\overline{d}_{it}^{*};S_{t},B_{t}  )=\E\Big[Y_{it}(\overline{d}_{it})-Y_{it}(\overline{d}_{it}^{*})|\overline{\mathbf{X}}_{it}\in S_{t}, \overline{D}_{i,k(t)}\in B_{t} \Big],  
\end{equation}
where $S_{t}$ is a subset of the support set of $\overline{\mathbf{X}}_{it}=\{\mathbf{X}_{ij}\}_{j=1}^{t}$, $\overline{D}_{i,k(t)}$ is a subsequence of $\overline{D}_{it}$, and $B_{t}$ is a subset of the support of $\overline{D}_{i,k(t)}$. Note that when $S_{t}$ is the support set of $\overline{\mathbf{X}}_{it}$,  $ATE_{t}(\overline{d}_{it},\overline{d}_{it}^{*};S_{t},B_{t})=ATE_{t}(\overline{d}_{it},\overline{d}_{it}^{*};B_{t})$.
If we would like to further restrict $B_{t}$ to be a singleton, say, $\overline{D}_{i,k(t)}=\overline{D}_{i,t-1}$ and $B_{t}=\{\overline{d}_{i,t-1}\}$, $ATE_{t}(\overline{d}_{it},\overline{d}_{it}^{*};S_{t},B_{t})$ is a generalization the usual concept of the average treatment effect on the treated (ATET). Similarly when $B_{t}$ is the support of $\overline{D}_{i,k(t)}$, $ATE_{t}(\overline{d}_{it},\overline{d}_{it}^{*};S_{t},B_{t})=ATE_{t}(\overline{d}_{it},\overline{d}_{it}^{*};S_{t})$. If we would like to further restrict $S_{t}$ to be a singleton, say, $S_{t}=\{\overline{\mathbf{x}}_{it}^{\dagger}\}$, $ATE_{t}(\overline{d}_{it},\overline{d}_{it}^{*};S_{t},B_{t})$ becomes the strata-specific average treatment effect. Finally, when $S_{t}$ is the support set of $\overline{\mathbf{X}}_{it}$ and $B_{t}$ is the support of $\overline{D}_{i,k(t)}$, $ATE_{t}(\overline{d}_{it},\overline{d}_{it}^{*};S_{t},B_{t})$ reduces to the overall average treatment effect $ATE_{t}(\overline{d}_{it},\overline{d}_{it}^{*})$.

So in the case of panel data with a continuous treatment, we have a lot of causal estimands to consider. Aggregated treatment effects are therefore needed for applications. \citet{callaway2024difference} call the causal estimands defined in different treatment levels and at different time periods, such as those in (\ref{estm1}), as the \textit{building blocks/components} for aggregate treatment effects. But what aggregation should we use? \citet{callaway2024difference} study the two-way fixed effects (TWFE) method for models with a continuous treatment and find that when we aggregate various ATET causal components, even under strong parallel trend assumptions, the TWFE method does not deliver unbiased estimator of the aggregated treatment effect unless we further assume treatment effect homogeneity. We therefore opt for an aggregation of various ATE causal components $ATE_{t}(\overline{d}_{it},\overline{d}_{it}^{*})$.\footnote{Results on the aggregation of various ATE causal components $ATE_{t}(\overline{d}_{it},\overline{d}_{it}^{*}, \overline{\mathbf{x}}_{t}^{\dagger})$ can be derived similarly, provided that we have enough observation for the strata $\overline{\mathbf{X}}_{it}=\{\overline{\mathbf{x}}_{it}^{\dagger}\}$.}

To implement the aggregation, we need to decide on the choice of the benchmark treatment $\overline{d}_{it}^{*}$. If there is a never-treated group among all individuals, we can set $\overline{d}_{it}^{*}=(0,\dots, 0)$, otherwise, a reasonable choice is $\overline{d}_{it}^{*}=(\E[D_{i1}],\dots, \E[D_{it}])$. For a given benchmark treatment  $\overline{d}_{it}^{*}$, $ATE_{t}(\overline{d}_{it},\overline{d}_{it}^{*})$ only depends on $\overline{d}_{it}$, so we can just write it as  $ATE_{t}(\overline{d}_{it})$.

An aggregated ATE at time $t$ is to average $ATE_{t}(\overline{d}_{it})$ according to relative frequency of $\overline{d}_{it}$, i.e.,
\[
ATEW_{t}=\E[ATE_{t}(\overline{D}_{it})],  \label{estm0}
\]
and an overall aggregated ATE is the time average of all available $ATEW_{t}$'s:
\[
ATEW^{*}=\frac{1}{T-k}\sum_{t=k+1}^{T}ATEW_{t},
\]
where $k$ is the smallest value such that $ATEW_{k+1}$ is allowed to be calculated according to the model of the potential outcomes.

Meanwhile, as argued by the literature \citep[][etc.]{angrist2000interpretation,callaway2024difference,chernozhukov2022automatic,klosin2022estimating}, when the treatment is continuous, it is natural to consider the average causal responses as the default causal estimands, as they measures the sensitivity of average potential outcome to a unit change in the treatment at the current period.  Specifically, we define\footnote{A more general definition of average causal response takes the form of 
\[
ACR_{t}(\overline{d}_{it})= \sum_{s\leq t}w_{s}\frac{\partial \E\big[Y_{it}(\overline{d}_{it})\big]}{\partial d_{s}},
\]
where $w_{s}\geq 0, s=1, \dots, t, \sum_{s\leq t}w_{s}=1$ are given weights to average different causal responses. For example, suppose $Y_{t}(\overline{d}_{it})$ only depends on $d_{i,t-1}$ and $d_{it}$, 
then we can define ACR as
\[
ACR_{t}(\overline{d}_{it})= \frac{1}{2}\left(\frac{\partial \E\big[Y_{it}(\overline{d}_{it})\big]}{\partial d_{i,t-1}}+\frac{\partial \E\big[Y_{it}(\overline{d}_{it})\big]}{\partial d_{it}}\right).
\]
}

\[
ACR_{t}(\overline{d}_{it})=\frac{\partial \E\big[Y_{it}(\overline{d}_{it})\big]}{\partial d_{it}}=\underset{h\rightarrow 0}{\lim} \frac{\E\big[Y_{it}(\overline{d}_{i,t-1}, d_{it}+h)\big]-\E\big[Y_{it}(\overline{d}_{i,t-1}, d_{it})\big]}{h}.  \label{estm2}
\]
Similarly, an aggregated ACR at time $t$ is to average $ACR_{t}(\overline{d}_{it})$ according to relative frequency of $\overline{d}_{it}$, i.e.,
\[
ACRW_{t}=\E[ACR_{t}(\overline{D}_{it})],  \label{estm3}
\]
and an overall ACR is the time average of all available $ACRW_{t}$'s:
\[
ACRW^{*}=\frac{1}{T-k}\sum_{t=k+1}^{T}ACRW_{t},
\]
where $k$ is the smallest value such that $ACRW_{k+1}$ is allowed to be calculated according to the model of the potential outcomes.

\subsection{Models for potential outcomes}
After defining the causal estimands, we now discuss their identification. To achieve this purpose, we need a model for the potential outcomes and a set of identifying assumptions.

A model for the potential outcomes stipulates characteristics of the probability distribution of the potential outcomes. As in \citet{sobel2012does}, we propose two candidate models. The first model is
\begin{equation}
  \label{FEC-SEI0}
  \E\Big[Y_{it}(\overline{d}_{it})|\overline{\mathbf{X}}_{it}, U_{i}, \overline{V}_{t}\Big]=\beta'\mathbf{X}_{it}+\tau(\overline{d}_{it}, Z_{it})
  +U_{i}+V_{t}, 
\end{equation}
where $\beta$ is the coefficient parameter for the covariates and $\tau(\overline{d}_{it}, Z_{it})$ is a known function of $\overline{d}_{it}$ and $Z_{it}$, which for simplicity is assume to be a subset of $\mathbf{X}_{it}$. For the time being we assume that $\tau(\overline{d}_{it}, Z_{it})$ is linear in parameters. We'll discuss specification of the functional forms of $\tau(\cdot)$ later. The model (\ref{FEC-SEI0}) can be equivalently expressed in the following regression form
\begin{equation}  
  \label{FEC-SEI1}
  \begin{aligned}
    &Y_{it}(\overline{d}_{it})=\beta'\mathbf{X}_{it}+\tau(\overline{d}_{it}, Z_{it})
    +U_{i}+V_{t}+ \varepsilon_{it}(\overline{d}_{it}),   \\
    &\E\Big[\varepsilon_{it}(\overline{d}_{it})|\overline{\mathbf{X}}_{it}, U_{i}, \overline{V}_{t} \Big]=0,  ~~\forall i, t
  \end{aligned}
\end{equation}
We call the model (\ref{FEC-SEI0}) or (\ref{FEC-SEI1}) a fixed effects causal model with sequentially mean independent errors (FEC-SEI).

The second model is
\begin{equation}
  \label{FEC-STI0}
  \E\Big[Y_{it}(\overline{d}_{it})|\overline{\mathbf{X}}_{iT}, U_{i}, \overline{V}_{T}\Big]=\beta'\mathbf{X}_{it}+\tau(\overline{d}_{it}, Z_{it})
  +U_{i}+V_{t},  
\end{equation}
which can be equivalently expressed in the following regression form
\begin{equation} 
  \label{FEC-STI1}
  \begin{aligned}
    &Y_{it}(\overline{d}_{it})=\beta'\mathbf{X}_{it}+\tau(\overline{d}_{it}, Z_{it})
    +U_{i}+V_{t}+ \varepsilon_{it}(\overline{d}_{it}),   \\
    &\E\Big[\varepsilon_{it}(\overline{d}_{it})|\overline{\mathbf{X}}_{iT}, U_{i}, \overline{V}_{T} \Big]=0.  ~~\forall i, t
  \end{aligned}
\end{equation}
We call the model (\ref{FEC-STI0}) or (\ref{FEC-STI1}) a fixed effects causal model with strictly mean independent errors (FEC-STI).

The FEC-STI model use stronger assumptions than the FEC-SEI model, and it's up to the researchers themselves to decide which one is more appropriate for their applications.

\subsection{Aggregate treatment effects}
Under either the FEC-STI or the FEC-SEI model, we have
\begin{equation}
  \label{FECm2}
  Y_{it}(\overline{d}_{it})-Y_{it}(\overline{d}_{it}^{*})=\tau(\overline{d}_{it}, Z_{it})-\tau(\overline{d}_{it}^{*}, Z_{it})
  + \varepsilon_{it}(\overline{d}_{it})-\varepsilon_{it}(\overline{d}_{it}^{*}),  
\end{equation}
hence
\[
  ATE_{t}(\overline{d}_{it},\overline{d}_{it}^{*} ;S_{t},B_{t} )=\E\Big[\tau(\overline{d}_{it}, Z_{it})-\tau(\overline{d}_{it}^{*}, Z_{it})
  + \varepsilon_{it}(\overline{d}_{it})-\varepsilon_{it}(\overline{d}_{it}^{*}) \mid \overline{\mathbf{X}}_{it}\in S_{t}, \overline{D}_{i,k(t)}\in B_{t} \Big]
\]

As argued above, it is appropriate for us to focus on the situation with $S_{t}$ being the support of $\overline{\mathbf{X}}_{it}$ and $B_{t}$ being the support of $\overline{D}_{i,k(t)}$, in which case the building blocks of our causal estimands are
\[
ATE_{t}(\overline{d}_{it})=\E\big[\tau(\overline{d}_{it}, Z_{it})-\tau(\overline{d}_{it}^{*}, Z_{it})\big]=\E\big[\tau(\overline{d}_{it}, Z_{it})\big]-\E\big[\tau(\overline{d}_{it}^{*}, Z_{it})\big]
\]
and
\[
  ACR_{t}(\overline{d}_{it})=\frac{\partial ATE_{t}(\overline{d}_{it})}{\partial d_{it}}=\frac{\partial \E\big[\tau(\overline{d}_{it}, Z_{it})\big]}{\partial d_{it}}
\]
The aggregated treatment effects $ATEW_{t}, ATEW^{*}, ACRW_{t}, ACRW^{*}$ can be then calculated with the building blocks of $ATE_{t}(\overline{d}_{it})$'s and $ACR_{t}(\overline{d}_{it})$'s according to the definitions given above.

\subsection{Identifying assumptions}
The FEC-STI model and the FEC-SEI model are models for the unobservable potential outcomes hence can not be used directly to derive estimators for the causal estimands. We need identifying assumptions to translate them into models for observable outcomes. As argued by \citet{sobel2012does}, in the case of panel data, we need to extend the classical conditional independence assumption (CIA) to its dynamic form to ensure the identifiability of the causal estimands. Specifically, we assume that
\begin{equation}
  \label{SCIA-I}
  \{\mathbf{X}_{ij}, V_{j}, Y_{ij}(\overline{d}_{ij}) \}_{j=t+1}^{T},Y_{it}(\overline{d}_{it}) \perp D_{it} \mid \overline{\mathbf{X}}_{it}, \overline{D}_{i,t-1}=\overline{d}_{i,t-1}, \overline{V}_{t}, U_i, ~\forall t=1,\dots, T.  
\end{equation}
The assumption in (\ref{SCIA-I}) is a set of sequentially conditional independence assumption (SCIA) and will be referred to as the SCIA-I assumption.
When the SCIA-I assumption holds, \citet{sobel2012does} shows that
\[
  \E\Big[Y_{it}(\overline{d}_{it}) \mid \overline{\mathbf{X}}_{it}, \overline{V}_{t}, U_i\Big]=\E\Big[Y_{it} \mid \overline{\mathbf{X}}_{it}, \overline{D}_{it}=\overline{d}_{it}, \overline{V}_{t}, U_i\Big].
\]
Therefore, with the FEC-SEI model,
\[
  \E\Big[Y_{it} \mid \overline{\mathbf{X}}_{it}, \overline{D}_{it}=\overline{d}_{it}, \overline{V}_{t}, U_{i}\Big]=\beta'\mathbf{X}_{it}+\tau(\overline{d}_{it}, Z_{it})
  +U_{i}+V_{t}.
\]
or equivalently
\[
  \E\Big[Y_{it} \mid \overline{\mathbf{X}}_{it}, \overline{D}_{it}, \overline{V}_{t}, U_{i}\Big]=\beta'\mathbf{X}_{it}+\tau(\overline{D}_{it}, Z_{it})
  +U_{i}+V_{t}.
\]
Define
\begin{equation}
  \label{resid}
  \varepsilon_{it}=Y_{it}-\big(\beta'\mathbf{X}_{it}+\tau(\overline{D}_{it}, Z_{it})
  +U_{i}+V_{t}\big),   
\end{equation}
we have
\begin{equation} 
  \label{PDM}
\begin{gathered}
  Y_{it}= \beta'\mathbf{X}_{it}+\tau(\overline{D}_{it}, Z_{it})+U_{i}+V_{t}+ \varepsilon_{it}, \\
  \E[\varepsilon_{it} \mid \overline{\mathbf{X}}_{it}, \overline{D}_{it}, \overline{V}_{t}, U_{i}] = 0.
\end{gathered}
\end{equation}
Equations (\ref{PDM}) describe a typical panel data regression model, where  the second equation in (\ref{PDM}) characterizes the property of its error terms. Specifically, we have
\[
  \E[\varepsilon_{it} \mid \mathbf{X}_{is}] =\E[\varepsilon_{it} \mid D_{is}]= \E[\varepsilon_{it} \mid V_{s}]=\E[\varepsilon_{it} \mid U_{i}]= 0, \forall t=1,\dots, T; s=1,\dots, t.
\]
Technically, (\ref{PDM}) specifies a TWFE regression model with pre-determined regressors, which is a special case of the so called dynamic panel data models with two-way fixed effects \citep{arellano1991some}. Parameter estimation for (\ref{PDM}) is usually done through generalized method of moments (GMM). 

An additional set of sequentially conditional independence assumption is
\begin{equation}
  \label{SCIA-II}
  Y_{it}(\overline{d}_{it}) \perp \overline{D}_{i\{j>t\}} \mid \overline{\mathbf{X}}_{iT}, \overline{D}_{it}=\overline{d}_{it}, \overline{V}_{T}, U_i, ~\forall t=1,\dots, T-1,  
\end{equation}
which together with the SCIA-I assumption will be referred to as the SCIA-II assumption. When the SCIA-II assumptions hold, we have that
\[
\E\Big[Y_{it} \mid \overline{\mathbf{X}}_{iT}, \overline{D}_{iT}=\overline{d}_{iT}, \overline{V}_{T}, U_i\Big] = \E\Big[Y_{it}(\overline{d}_{it}) \mid \overline{\mathbf{X}}_{iT}, \overline{V}_{T}, U_i\Big].
\]
Hence with the FEC-STI model, we have that
\[
  \E\Big[Y_{it} \mid \overline{\mathbf{X}}_{iT}, \overline{D}_{iT}, \overline{V}_{T}, U_{i}\Big]=\beta'\mathbf{X}_{it}+\tau(\overline{D}_{it}, Z_{it})
  +U_{i}+V_{t}.
\]
Let's define $\varepsilon_{it}$ as in (\ref{resid}), we have
\begin{equation} 
  \label{STM}
  \begin{gathered}
  Y_{it}= \beta'\mathbf{X}_{it}+\tau(\overline{D}_{it}, Z_{it})
  +U_{i}+V_{t}+ \varepsilon_{it} \\
  \E[\varepsilon_{it} \mid \overline{\mathbf{X}}_{iT}, \overline{D}_{iT}, \overline{V}_{T}, U_{i}] = 0.
  \end{gathered}
\end{equation}
The equations (\ref{STM}) define a TWFE model with strictly exogenous regressors, and its parameters can be estimated with typical methods such as the LSDV.

Note that time fixed effects $V_{t}$'s in our framework are random variables. In typical TWFE regressions, individual effects $U_{i}$'s are assumed to be random variables having arbitrary correlation with the regressors, while time effects are included in the regression equation as dummy variables \citep[see][]{arellano1991some,roodman2009xtabond2,chernozhukov2017}, which implicitly assumes that $V_{t}$'s are nonrandom variables. Our models and assumptions can be simplified if we follow that tradition.

\section{Estimation of aggregate treatment effects}
In this section, we show how to build estimators for the causal estimands defined in Section 2.2. It can be seen from the derivation in Section 2.4 that under whichever model, causal estimands are determined by the concrete form of $\tau(\overline{d}_{it}, Z_{it})$. Remember that we have assume that $\tau(\overline{d}_{it}, Z_{it})$ is linear in its parameters, so without loss of generality, we suppose that $\tau(\overline{d}_{it}, Z_{it})$ has the following form\footnote{Specific examples are discussed in Appendix A.}  
\[
\tau(\overline{D}_{it},Z_{it}) = \tau_1 M_{it1} + \tau_2 M_{it2} + \cdots + \tau_S M_{itS},
\]
where $\tau_1$, $\ldots$, $\tau_S$ are the unknown parameters and $M_{it1}$, $\ldots$, $M_{itS}$ are $S$ known functions of $\overline{d}_{it}$ and $Z_{it}$. Then, it is obvious that if we obtain estimators for the parameters $\tau_1,\ldots,\tau_S$, we can plug them into the expression of causal estimands to build plug-in estimators for causal estimands. Therefore, let $\phi = (\beta^\prime,\tau_1,\ldots,\tau_S)^\prime$ denote all the parameters to be estimated, we first address the estimation problem of $\phi$. 

We commence with how to estimate $\phi$ in the model FEC-SEI. To avoid the difficulties caused by the unknown fixed effects in the estimation, we first take the first difference of \eqref{PDM} to eliminate the individual fixed effects,
\begin{equation}
\label{fd}
\Delta Y_{it}= \beta^\prime \Delta \mathbf{X}_{it} + \tau_1 \Delta M_{it1} + \tau_2 \Delta M_{it2} + \cdots + \tau_S \Delta M_{itS}+ \Delta V_{t}+ \Delta \varepsilon_{it},
\end{equation}
where $\Delta Y_{it} = Y_{it} - Y_{i,t-1}$, and the remaining variables are defined similarly. 

Then we define the cross-sectional mean as 
\begin{equation}
\label{cm}
\Delta Y_{t} = \Delta Y_{it} - \frac{1}{N} \sum_{i=1}^N \Delta Y_{it} = \beta^\prime \Delta \mathbf{X}_{t} + \tau_1 \Delta M_{i1} + \tau_2 \Delta M_{i2} + \cdots + \tau_S \Delta M_{iS}+ \Delta V_{t}+ \Delta \varepsilon_{t},
\end{equation}
where $\Delta \mathbf{X}_t$, $\Delta M_{i1}$, \ldots, $\Delta M_{iS}$, $\Delta \varepsilon_t$ are similarly defined as $\Delta Y_t$.

By subtracting \eqref{cm} from \eqref{fd}, we can further eliminate time fixed effects to obtain
\[
\Delta Y_{it}^* = \beta^\prime \Delta \mathbf{X}_{it}^* + \tau_1 \Delta M_{it1}^* + \tau_2 \Delta M_{it2}^* + \cdots + \tau_S \Delta M_{itS}^* + \Delta \varepsilon_{it}^*,
\]
where $\Delta Y_{it}^* = \Delta Y_{it} - \Delta Y_t$, and other variables are defined similarly. 

Note that according to the property of the errors in \eqref{PDM}, we have $\E[\varepsilon_{it} \mid \overline{\mathbf{X}}_{i,t-1}, \overline{D}_{i,t-1}] = 0$ and $\E[\varepsilon_{i,t-1} \mid \overline{\mathbf{X}}_{i,t-1}, \overline{D}_{i,t-1}] = 0$, which imply $\E[\Delta \varepsilon_{it} \mid \overline{\mathbf{X}}_{i,t-1}, \overline{D}_{i,t-1}] = 0$. Because we have assumed that variables of different individuals are i.i.d., the last equation yields 
\[
\E[\Delta \varepsilon_{it}^* \mid \overline{\mathbf{X}}_{i,t-1}, \overline{D}_{i,t-1}] = 0,
\]
which implies the following moment conditions for $i=1,\ldots,N$ and $t=2,\ldots,T$,
\[
E\Big[\big(\mathbf{X}_{i1}^\prime,\ldots,\mathbf{X}_{i,t-1}^\prime,D_{i1},\ldots,D_{i,t-1}\big)^\prime \big(\Delta Y_{it}^* - \beta^\prime \Delta \mathbf{X}_{it}^* - \tau_1 \Delta M_{it1}^* - \cdots - \tau_S \Delta M_{itS}^*\big)\Big]=\mathbf{0}.
\]

Define the instrument matrix for individual $i$ as 
\[
\mathbf{W}_i = \begin{bmatrix}
    (\mathbf{X}_{i1}^\prime,D_{i1}) & \mathbf{0}^\prime & \cdots & \mathbf{0}^\prime \\
    \mathbf{0}^\prime & (\overline{\mathbf{X}}_{i2}^\prime,\overline{D}_{i2}^\prime) & \cdots & \mathbf{0}^\prime \\
    \vdots & \vdots & \ddots & \vdots \\
    \mathbf{0}^\prime & \mathbf{0}^\prime & \cdots & (\overline{\mathbf{X}}_{i,T-1}^\prime,\overline{D}_{i,T-1}^\prime)
\end{bmatrix},
\]
and define the vector $\Delta \mathbf{Y}_i^*=(\Delta Y_{i2}^*,\ldots,\Delta Y_{iT}^*)^\prime$ and the design matrix 
\[
\mathbf{R}_i^* = \begin{bmatrix}
    \Delta \mathbf{X}_{i2}^{*\prime} & \Delta M_{i21}^* & \Delta M_{i22}^* & \cdots & \Delta M_{i2S}^* \\
    \Delta \mathbf{X}_{i3}^{*\prime} & \Delta M_{i31}^* & \Delta M_{i32}^* & \cdots & \Delta M_{i3S}^* \\
    \vdots & \vdots & \vdots & \ddots & \vdots \\
    \Delta \mathbf{X}_{iT}^{*\prime} & \Delta M_{iT1}^* & \Delta M_{iT2}^* & \cdots & \Delta M_{iTS}^*
\end{bmatrix}.
\]
Then the moment conditions can be rewritten in the matrix form 
\[
\E[\mathbf{W}_i^\prime(\Delta \mathbf{Y}_i^*-\mathbf{R}_i^*\phi)]=\mathbf{0}.
\]

As mentioned before, we will use the GMM to estimate $\phi$. To be specific, the GMM estimator of $\phi$ is obtained by solving the optimization problem 
\[
\widehat{\phi}^G = \mathop{\arg\min}_\phi \Big(\frac{1}{N}\sum_{i=1}^N \mathbf{W}_i^\prime(\Delta \mathbf{Y}_i^*-\mathbf{R}_i^*\phi)\Big)^\prime \boldsymbol{\Omega} \Big(\frac{1}{N}\sum_{i=1}^N \mathbf{W}_i^\prime(\Delta \mathbf{Y}_i^*-\mathbf{R}_i^*\phi)\Big),
\]
where $\boldsymbol{\Omega}$ is a positive definite weighting matrix and $\widehat{\phi}^G$ has the following solution 
\[
\widehat{\phi}^G = \bigg[\bigg(\frac{1}{N}\sum_{i=1}^N \mathbf{W}_i^\prime \mathbf{R}_i^*\bigg)^\prime \boldsymbol{\Omega} \bigg(\frac{1}{N}\sum_{i=1}^N \mathbf{W}_i^\prime \mathbf{R}_i^*\bigg) \bigg]^{-1} \bigg[\bigg(\frac{1}{N}\sum_{i=1}^N \mathbf{W}_i^\prime \mathbf{R}_i^*\bigg)^\prime \boldsymbol{\Omega} \bigg(\frac{1}{N}\sum_{i=1}^N \mathbf{W}_i^\prime \Delta \mathbf{Y}_i^* \bigg)\bigg].
\]

Next, we show how to estimate $\phi$ under the model FEC-STI. Same as before, we can eliminate the unobserved fixed effects to avoid the difficulties in the estimation. Following the common practice in the literature, when the errors are strictly mean independent of the regressors, we may use the two-way within transformation to eliminate the fixed effects. Specifically, for a general variable $V_{it}$, we define its two-way within transformed counterpart as 
\[
\ddot{V}_{it} = V_{it} - V_{i\cdot} - V_{\cdot t} + V_{\cdot\cdot},
\]
where $V_{i\cdot} = \frac{1}{T}\sum_{t=1}^T V_{it}$, $V_{\cdot t} = \frac{1}{N}\sum_{i=1}^N V_{it}$, and $V_{\cdot\cdot} = \frac{1}{NT}\sum_{i=1}^N \sum_{t=1}^T V_{it}$. Then, we apply this transformation to each variable in FEC-STI, and we would obtain a transformed model without individual and time fixed effects,
\begin{equation}
\label{sti}
\ddot{Y}_{it} = \beta^\prime \ddot{\mathbf{X}}_{it} + \tau_1 \ddot{M}_{it1} + \tau_2 \ddot{M}_{it2} + \cdots + \tau_S \ddot{M}_{itS} + \ddot{\varepsilon}_{it}.
\end{equation}
Note that according to the properties of the errors shown in \eqref{STM}, we have $\E[\varepsilon_{it} \mid \overline{\mathbf{X}}_{iT}, \overline{D}_{iT}] = 0, \forall t$, which implies $\E[\varepsilon_{it} - \frac{1}{T}\sum_{t=1}^T \varepsilon_{it} \mid \overline{\mathbf{X}}_{iT}, \overline{D}_{iT}] = 0$. Because we have assumed that variables of different individuals are i.i.d., we further have $\E[\varepsilon_{it} - \frac{1}{T}\sum_{t=1}^T \varepsilon_{it} \mid \overline{\mathbf{X}}_{1T}, \overline{D}_{1T}, \ldots,\allowbreak \overline{\mathbf{X}}_{NT}, \overline{D}_{NT}] = 0$,
which suggests 
\begin{align*}
& \E\left[\varepsilon_{it} - \frac{1}{T}\sum_{t=1}^T \varepsilon_{it} - \frac{1}{N}\sum_{i=1}^N\left(\varepsilon_{it} - \frac{1}{T}\sum_{t=1}^T \varepsilon_{it}\right) ~ \bigg| ~ \overline{\mathbf{X}}_{1T}, \overline{D}_{1T}, \ldots, \overline{\mathbf{X}}_{NT}, \overline{D}_{NT}\right] \\
& = \E[\ddot{\varepsilon}_{it} \mid \overline{\mathbf{X}}_{1T}, \overline{D}_{1T}, \ldots, \overline{\mathbf{X}}_{NT}, \overline{D}_{NT}]= 0, \quad \forall i, t.
\end{align*}
Note that $\ddot{\mathbf{X}}_{it}$, $\ddot{M}_{it1}$, $\ddot{M}_{it2}$, \ldots, $\ddot{M}_{itS}$ are all functions of $\overline{\mathbf{X}}_{1T}, \overline{D}_{1T}, \ldots, \overline{\mathbf{X}}_{NT}, \overline{D}_{NT}$, thus regressors in \eqref{sti} are strictly exogenous and we can use OLS to obtain a TWFE estimator of $\phi$.

Define the design matrix of individual $i$,
\[
\ddot{\mathbf{D}}_i = \begin{bmatrix}
    \ddot{\mathbf{X}}_{i1}^\prime & \ddot{M}_{i11} & \ddot{M}_{i12} & \cdots & \ddot{M}_{i1S} \\
    \ddot{\mathbf{X}}_{i2}^\prime & \ddot{M}_{i21} & \ddot{M}_{i22} & \cdots & \ddot{M}_{i2S} \\
    \vdots & \vdots & \ddots & \vdots \\
    \ddot{\mathbf{X}}_{iT}^\prime & \ddot{M}_{iT1} & \ddot{M}_{iT2} & \cdots & \ddot{M}_{iTS} 
\end{bmatrix},
\]
and the vector $\ddot{\mathbf{Y}}_i = (\ddot{Y}_{i1},\ldots,\ddot{Y}_{iT})^\prime$. The TWFE estimator $\widehat{\phi}^O$ has the solution 
\[
\widehat{\phi}^O = \left(\frac{1}{N}\sum_{i=1}^N \ddot{\mathbf{D}}_i^\prime \ddot{\mathbf{D}}_i\right)^{-1}\left(\frac{1}{N}\sum_{i=1}^N \ddot{\mathbf{D}}_i^\prime \ddot{\mathbf{Y}}_i\right).
\]

With the estimator $\widehat{\phi}^G$ or $\widehat{\phi}^O$ in hand, we then demonstrate the construction of plug-in estimators for causal estimands under different specific choices of $\tau(\overline{d}_{it}, Z_{it})$. The principle is very simple, that is, to replace the parameters and expectations with the corresponding parameter estimators and sample averages, respectively. For brevity, here we will only present the estimation of $ACR_{t}(\overline{d}_{it})$ and $ACRW^{*}$ under $\tau(\overline{d}_{it}, Z_{it})=\tau_{1} d_{it}+\tau_{2} d_{it}^{2}+ \tau_{3} d_{it}Z_{it}$, with more examples given in the Appendix.

Note that under this choice, we have  
\[
  ACR_{t}(\overline{d}_{it})=\frac{\partial \E\big[\tau(\overline{d}_{it}, Z_{it})\big]}{\partial d_{it}} =\tau_{1} +2\tau_{2} d_{it} +\tau_{3} \E[Z_{it}],
\]
and
\[
  \begin{aligned}
    ACRW_{t}&=\tau_{1} +2\tau_{2} \E[D_{it}]+\tau_{3} \E[Z_{it}], \\
    ACRW^{*}&=\frac{1}{T}\sum_{t=1}^{T}ACRW_{t}  =\tau_{1} +2\tau_{2} \frac{1}{T}\sum_{t=1}^{T}\E[D_{it}] + \tau_{3}\frac{1}{T}\sum_{t=1}^{T}\E[Z_{it}]
  \end{aligned}
\]
Let $\hat{\tau}_{1}, \hat{\tau}_{2}, \hat{\tau}_{3}$ be the (GMM or TWFE) estimators for $\tau_{1}, \tau_{2}, \tau_{3}$, respectively. A consistent estimator for $ACRW_{t}$ is
\[
\begin{aligned}
\widehat{ACRW}_{t}&=\hat{\tau}_{1} +2\hat{\tau}_{2}\hat{\E}[D_{it}] +\hat{\tau}_{3}\hat{\E}[Z_{it}]\\
&=\hat{\tau}_{1} +2\hat{\tau}_{2}\frac{1}{N}\sum_{i=1}^{N}D_{it} + \hat{\tau}_{3}\frac{1}{N}\sum_{i=1}^{N}Z_{it}
\end{aligned}
\]
and a consistent estimator for $ACRW^{*}$ is
\[
  \widehat{ACRW}^{*}=\hat{\tau}_{1} +2\hat{\tau}_{2}\frac{1}{NT}\sum_{t=1}^{T}\sum_{i=1}^{N}D_{it} + \hat{\tau}_{3}\frac{1}{NT}\sum_{t=1}^{T}\sum_{i=1}^{N}Z_{it}.
\]

\section{Asymptotic theory}
In this section, we show that our estimators defined in the last section and Appendix A for causal estimands are asymptotically normal. This enables us to conduct statistical inference for the causal effects of interest in a standard manner. 

We first develop the asymptotic theory for the estimators of causal estimands under the model FEC-SEI. According to the standard GMM theory, $\widehat{\phi}^G$ is asymptotically normal. Observing that $\widehat{ATE}_t$ and $\widehat{ACR}_t$ are functions of $\widehat{\phi}^G$, it follows from the delta method that $\widehat{ATE}_t$ and $\widehat{ACR}_t$ are also asymptotically normal. However, $\widehat{ATEW}_t$, $\widehat{ATEW}^*$, $\widehat{ACRW}_t$, $\widehat{ACRW}^*$ are functions of $\widehat{\phi}^G$ and some sample averages, so asymptotic normality of $\widehat{\phi}^G$ alone does not suffice to deduce their asymptotic normality. Therefore, we subsequently propose a theorem to show that they (actually this theorem applies to $\widehat{ATE}_t$ and $\widehat{ACR}_t$ as well) are also asymptotically normal under common regularity conditions for GMM. 

Let $CE$ denote any causal estimand defined in Section 3 and Appendix A, and let $\widehat{CE}$ denote the corresponding plug-in estimator defined in Section 3 and Appendix A. If we use $\mathbf{Q}_i$ to denote the sample averages that appear in $\widehat{CE}$\footnote{For example, under $\tau(\overline{d}_{it}, Z_{it})=\tau_{1} d_{it}+\tau_{2} d_{it}^{2}$, $\mathbf{Q}_i = D_{it}$ for $\widehat{ACRW}_t$ and $\mathbf{Q}_i = \frac{1}{T}\sum_{t=1}^T D_{it}$ for $\widehat{ACRW}^*$. Under $\tau(\overline{d}_{it}, Z_{it})=\tau_{1} d_{it}+\tau_{2} d_{it}^{2}+ \tau_{3} d_{it}Z_{it}$, $\mathbf{Q}_i = (D_{it}, Z_{it})^\prime$ for $\widehat{ACRT}_t$ and $\mathbf{Q}_i = (\frac{1}{T}\sum_{t=1}^T D_{it}, \frac{1}{T}\sum_{t=1}^T Z_{it})^\prime$ for $\widehat{ACRW}^*$.}, it is easy to observe that for each causal estimand, there exists a function $f(\cdot)$ such that $CE=f(\phi,\E(\mathbf{Q}_i))$ and $\widehat{CE} = f(\widehat{\phi}^G, \frac{1}{N}\sum_{i=1}^N \mathbf{Q}_i)$. Then, we define 
\[
\mathbf{R}_i = \begin{bmatrix}
    \Delta \mathbf{X}_{i2}^{\prime} & \Delta M_{i21} & \cdots & \Delta M_{i2S} \\
    \vdots & \vdots & \ddots & \vdots \\
    \Delta \mathbf{X}_{iT}^{\prime} & \Delta M_{iT1} & \cdots & \Delta M_{iTS}
\end{bmatrix},
\]
$\mathbf{G} = \E(\mathbf{W}_i^\prime\mathbf{R}_i) - \E(\mathbf{W}_i^\prime)\E(\mathbf{R}_i)$, and $\boldsymbol{\Sigma} = \mathrm{Var}[((\mathbf{W}_i^\prime \Delta \boldsymbol{\varepsilon}_i)^\prime, \mathrm{Vec}(\mathbf{W}_i^\prime)^\prime, \Delta \boldsymbol{\varepsilon}_i^\prime, \mathbf{Q}_i^\prime)^\prime]$. It is easy to see that for vectors $w,x,y,z$, we can define a function $h(w,x,y,z) = (w - \mathrm{Mat}(x)y, z)^\prime$ such that 
\[
h\left(\begin{bmatrix}
    \frac{1}{N} \sum_{i=1}^N \mathbf{W}_i^\prime \Delta \boldsymbol{\varepsilon}_i\\
    \frac{1}{N} \sum_{i=1}^N \mathrm{Vec}(\mathbf{W}_i^\prime) \\
    \frac{1}{N} \sum_{i=1}^N \Delta \boldsymbol{\varepsilon}_i\\
    \frac{1}{N} \sum_{i=1}^N \mathbf{Q}_i 
\end{bmatrix}\right) = \begin{bmatrix}
    \frac{1}{N}\sum_{i=1}^N \mathbf{W}_i^\prime \Delta \boldsymbol{\varepsilon}_i - \frac{1}{N}\sum_{i=1}^N \mathbf{W}_i^\prime \frac{1}{N}\sum_{i=1}^N \Delta \boldsymbol{\varepsilon}_i \\
    \frac{1}{N} \sum_{i=1}^N \mathbf{Q}_i
\end{bmatrix},
\]
where $\mathrm{Mat}(x)$ organizes the vector $x$ into a matrix of a suitable shape. Denote 
\[
\mathbf{B} = \frac{\partial h(w,x,y,z)}{\partial (w,x,y,z)^\prime}\bigg|_{(x,w,y,z)^\prime=(\mathbf{0}^\prime,\E[\mathrm{Vec}(\mathbf{W}_i^\prime)]^\prime,\mathbf{0}^\prime,\E(\mathbf{Q}_i)^\prime)^\prime}, \quad \mathbf{H} = \begin{bmatrix}
    (\mathbf{G}^\prime\boldsymbol{\Omega}\mathbf{G})^{-1}\mathbf{G}^\prime\boldsymbol{\Omega} & \mathbf{0} \\
    \mathbf{0} & \mathbf{I}
\end{bmatrix},
\]
where we assume that $\mathbf{G}^\prime\boldsymbol{\Omega}\mathbf{G}$ is invertible, and define $\nabla_f(\phi,\E(\mathbf{Q}_i))$ as the gradient of $f(\cdot)$ evaluated at $(\phi,\E(\mathbf{Q}_i))$. The asymptotic result is stated in the following theorem. 
\begin{theorem}
\label{th1}
Under the model FEC-SEI, SCIA-I, and standard regularity conditions for GMM estimators (e.g., see the conditions of Theorem 14.2 in \citet{wooldridge2010econometric}), we have 
\[
\sqrt{N} (\widehat{CE} - CE) \dto N\Big(0,\nabla_f(\phi,\E(\mathbf{Q}_i))^\prime\mathbf{H}\mathbf{B}\boldsymbol{\Sigma}\mathbf{B}^\prime\mathbf{H}^\prime\nabla_f(\phi,\E(\mathbf{Q}_i))\Big),
\]
where we assume that $\mathbf{B}\boldsymbol{\Sigma}\mathbf{B}^\prime$ is positive definite and $\nabla_f(\phi,\E(\mathbf{Q}_i))^\prime\mathbf{H}\mathbf{B}\boldsymbol{\Sigma}\mathbf{B}^\prime\mathbf{H}^\prime\nabla_f(\phi,\E(\mathbf{Q}_i)) \allowbreak > 0$.
\end{theorem}

Next, we turn our attention to the asymptotic normality of the estimators for causal estimands under the model FEC-STI. We continue to denote $\mathbf{Q}_i$ as the sample averages that appear in $\widehat{CE}$, and there exists $f(\cdot)$ that is the same as before such that $CE=f(\phi,\E(\mathbf{Q}_i))$ and $\widehat{CE} = f(\widehat{\phi}^O, \frac{1}{N}\sum_{i=1}^N \mathbf{Q}_i)$. For a general variable $V_{it}$, we define $\dot{V}_{it} = V_{it} - \frac{1}{T}\sum_{t=1}^T V_{it}$, so $\ddot{V}_{it} = \dot{V}_{it} - \frac{1}{N} \sum_{i=1}^N \dot{V}_{it}$. We also define 
\[
\dot{\mathbf{D}}_i = \begin{bmatrix}
    \dot{\mathbf{X}}_{i1}^\prime & \dot{M}_{i11} & \dot{M}_{i12} & \cdots & \dot{M}_{i1S} \\
    \dot{\mathbf{X}}_{i2}^\prime & \dot{M}_{i21} & \dot{M}_{i22} & \cdots & \dot{M}_{i2S} \\
    \vdots & \vdots & \ddots & \vdots \\
    \dot{\mathbf{X}}_{iT}^\prime & \dot{M}_{iT1} & \dot{M}_{iT2} & \cdots & \dot{M}_{iTS} 
\end{bmatrix},
\]
$\mathbf{U} = \E(\dot{\mathbf{D}}_i^\prime \dot{\mathbf{D}}_i) - \E(\dot{\mathbf{D}}_i^\prime)\E(\dot{\mathbf{D}}_i)$, and $\boldsymbol{\Xi} = \mathrm{Var}[((\dot{\mathbf{D}}_i^\prime \dot{\boldsymbol{\varepsilon}}_i)^\prime,\mathrm{Vec}(\dot{\mathbf{D}}_i^\prime)^\prime,\dot{\boldsymbol{\varepsilon}}_i^\prime,\mathbf{Q}_i^\prime)^\prime]$. It is easy to see that for vectors $w,x,y,z$, we can define a function $g(w,x,y,z) = (w-\mathrm{Mat}(x)y,z)^\prime$ such that 
\[
g\left(\begin{bmatrix}
    \frac{1}{N} \sum_{i=1}^N \dot{\mathbf{D}}_i^\prime \dot{\boldsymbol{\varepsilon}}_i\\
    \frac{1}{N} \sum_{i=1}^N \mathrm{Vec}(\dot{\mathbf{D}}_i^\prime) \\
    \frac{1}{N} \sum_{i=1}^N \dot{\boldsymbol{\varepsilon}}_i\\
    \frac{1}{N} \sum_{i=1}^N \mathbf{Q}_i
\end{bmatrix}\right) = \begin{bmatrix}
    \frac{1}{N}\sum_{i=1}^N \dot{\mathbf{D}}_i^\prime \dot{\boldsymbol{\varepsilon}}_i - \frac{1}{N}\sum_{i=1}^N \dot{\mathbf{D}}_i^\prime\frac{1}{N}\sum_{i=1}^N \dot{\boldsymbol{\varepsilon}}_i \\
    \frac{1}{N} \sum_{i=1}^N \mathbf{Q}_i
\end{bmatrix}.
\]
Define 
\[
\mathbf{C} = \frac{\partial g(w,x,y,z)}{\partial (w,x,y,z)^\prime}\bigg|_{(x,w,y,z)^\prime=(\mathbf{0}^\prime,\E[\mathrm{Vec}(\dot{\mathbf{D}}_i^\prime)]^\prime,\mathbf{0}^\prime,\E(\mathbf{Q}_i)^\prime)^\prime}, \quad \mathbf{J} = \begin{bmatrix}
    \mathbf{U}^{-1} & \mathbf{0} \\
    \mathbf{0} & \mathbf{I}
\end{bmatrix},
\]
where we assume that $\mathbf{U}$ is invertible, and once again define $\nabla_f(\phi,\E(\mathbf{Q}_i))$ as the gradient of $f(\cdot)$ evaluated at $(\phi,\E(\mathbf{Q}_i))$. We formally state the asymptotic result as follows.
\begin{theorem}
\label{th2}
Under the model FEC-STI, SCIA-II, we have 
\[
\sqrt{N}(\widehat{CE}-CE) \dto N\Big(0,\nabla_f(\phi,\E(\mathbf{Q}_i))^\prime\mathbf{J}\mathbf{C}\boldsymbol{\Xi}\mathbf{C}^\prime\mathbf{J}^\prime\nabla_f(\phi,\E(\mathbf{Q}_i))\Big),
\]
where we assume that $\mathbf{C}\boldsymbol{\Xi}\mathbf{C}^\prime$ is positive definite and $\nabla_f(\phi,\E(\mathbf{Q}_i))^\prime\mathbf{J}\mathbf{C}\boldsymbol{\Xi}\mathbf{C}^\prime\mathbf{J}^\prime\nabla_f(\phi,\E(\mathbf{Q}_i)) > 0$.
\end{theorem}

By Theorems 1 and 2, the causal estimators have asymptotically normal distribution. Though we have provided analytical asymptotic variances for the estimators, in pratice we recommend using the cross-sectional resampling method \citep{kapetanios2008bootstrap} to obtain the variances and standard errors of $\widehat{ACRW}_{t}$ and $\widehat{ACRW}^{*}$. Specifically, in each resampling, we randomly draw with replacement an index set $(i_{1},i_{2},\dots,i_{N})$ from $(1,2,\dots,N)$, and obtain all the time series observations of the $(i_{1},i_{2},\dots,i_{N})$ as the bootstrap sample. \citet{chernozhukov2017} also recommend such cross-sectional bootstrap for inference with panel data, and their simulations indicate that the bootstrap standard errors are more stable and reliable than the analytical standard errors across different modeling assumptions and estimation methods. Moreover, the nonparametric bootstrap is usually much easier to implement with standard statistical software packages as compared with the analytical methods.

\section{An application}
In this part we illustrate how to apply our proposed method in an toy example. Our data come from \citet{SP3/YW9OWQ_2022}.\footnote{The original problem, i.e., the causal effect of foreign aid and economic growth, was stated in \citet{dreher2020aid}, where foreign was treated as endogenous and instrumental variables were constructed to identify the causal effect of foreign aid on economic growth. As pointed out by \citet{dreher2020aid}, there is a long list of articles on the aid's effect on growth, with little consensus. It is not the purpose of this application to joint the debate on the final truth regrading to this issue. Here in this paper we just want to use this open dataset to illustrate how various causal effects can be defined and estimated with our proposed method. We also don't discuss how to use DAGs to validate/invalidate the sequentially conditional independent assumptions in this example as DAGs are just graphical representation of contextual knowledge and are better discussed in a separate applied paper.} In this problem, the outcome variable $Growth$ is growth rate of real GDP per capita, the treatment variable $Aid$ is foreign aid (over real GDP), which is continuous, and the vector of control variables includes logarithm of the initial real GDP per capital ($X_{1}$), assassinations ($X_{2}$), the interaction between ethic fractionalization and assassinations ($X_{3}$), and the one year lag of the ratio of broad money (M2) over real GDP ($X_{4}$).

We consider three alternative specifications for $\tau()$: (1) $\tau(Aid,X_{1})=\beta_{1}Aid$, (2) $\tau(Aid,X_{1})=\beta_{1}Aid + \beta_{2}Aid^{2}$, and (3) $\tau(Aid,X_{1})=\beta_{1}Aid + \beta_{2}Aid^{2} + \beta_{3}Aid\cdot X_{1}$. Our main causal estimand is $ACRW^{*}$, and we also consider causal estimands $ACRW_{t}, t=1,...,T$.

Let's first assume a FEC-STI model along with the SCIA-II assumption. We then have the following three alternative econometrics models for the observed data:
\begin{eqnarray*}
\text{Model 1:}~~Growth_{it} &=& \beta_{1}Aid_{i,t-1} + \mathbf{X}_{it}'\gamma + \eta_{i} + \tau_{t} + \varepsilon_{it}, \\
\text{Model 2:}~~Growth_{it} &=& \beta_{1}Aid_{i,t-1} + \beta_{2}Aid_{i,t-1}^{2}+ \mathbf{X}_{it}'\gamma + \eta_{i} + \tau_{t} + \varepsilon_{it}, \\
\text{Model 3:}~~Growth_{it} &=& \beta_{1}Aid_{i,t-1} + \beta_{2}Aid_{i,t-1}^{2} + \beta_{3}Aid_{i,t-1}\cdot X_{1i,t} + \mathbf{X}_{it}'\gamma \\
&+& \eta_{i} + \tau_{t} + \varepsilon_{it}, 
\end{eqnarray*}
where $i$ represents country and $t$ represents period (four-years),  $Growth_{it}$ is the average real per capita GDP growth rate of country $i$ over period $t$, and definitions of other variables are similar, $\eta_{i} $ and $\tau_{t}$ represent country fixed effects and period fixed effects, respectively, and $\varepsilon_{it}$ denotes random error. Under a FEC-STI model with the SCIA-II assumption, for any $t=1,\dots, T$, $\varepsilon_{it}$'s are conditionally mean independent of the right-hand side variables at all periods, hence the coefficients in Models 1-3 can be estimated by the TWFE method. After the coefficients are estimated, we can obtain estimates of $ACRW_{t}, t=1,...,T$ and $ACRW^{*}$ from the estimators proposed in Section 3, and obtain their standard errors by nonparametric panel bootstrap. 

We use part of the data that were behind columns (1)-(2) of Table 2 in Dreher and Langlotz (2020) to conduct our causal effects estimation. The data were unbalanced panel data with 97 units and 10 time periods. \footnote{Note that Models 1-2 are the models used by Dreher and Langlotz (2020) in producing results with TWFE in columns (1)-(2), respectively, of Table 2. We introduce in Model 3 an interaction between the treatment and an control variable to allow for more flexible form of treatment effects heterogeneity.} The summary statistics are reported in Table 1.
\begin{table}[htbp]
  \centering 
  \caption{Summary Statistics for Variables Used} 
\begin{tabular}{lccccc} 
\hline\hline
Variable & $N$ & Mean & Std. Dev. & Min & Max \\ 
\hline
GDPPC growth rate & 832 & 1.725 & 3.730 & $-$32.425 & 17.054 \\ 
Aid & 832 & 3.496 & 4.819 & $-$0.149 & 47.909 \\ 
Log(initialGDPPC) & 822 & 7.591 & 1.241 & 4.825 & 10.805 \\ 
Assassinations & 829 & 0.292 & 0.979 & 0.000 & 11.500 \\ 
Ethnic$\times$Assassinations & 829 & 0.109 & 0.422 & 0.000 & 7.360 \\ 
Lag of M2/GDP  & 813 & 36.791 & 28.085 & 0.357 & 238.304 \\ 
\hline\hline
\end{tabular} 
\end{table} 

\begin{figure}
\centering
\includegraphics[scale=0.9]{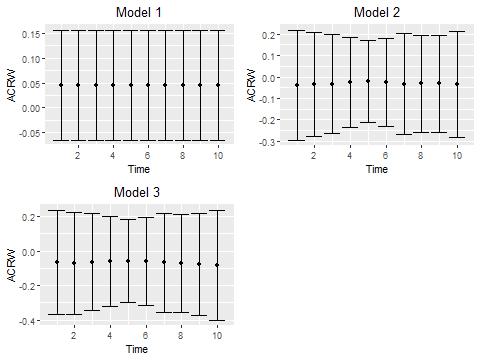}
\caption{Time paths of ACRW for Models 1-3 under FEC-STI and SCIA-II}
\label{fig: fig1}
\end{figure}

The estimation results are displayed in Table 2. Model 1 is a homogeneous treatment effects model in which 
\begin{equation}
ACRW(1)=\cdots = ACRW(T)=ACRW^{*}=\beta_{1}.
\end{equation}
I.e., under Model 1, the treatment effect is just $\beta_{1}$, and its estimate from data is 0.045 with a standard error of 0.057. In Models 2-3, treatment effects change with treatment level and/or level of the initial real GDP per capital; therefore we have many definitions of causal effects, and as we proposed in this paper, we use $ACRW(t)$'s and $ACRW^{*}$ as our causal estimands. The estimates of $ACRW^{*}$ for Models 2-3 are -0.031 and -0.068, respectively, and their bootstrap standard errors are 0.116 and 0.144, respectively. The $ACRW(t)$'s for Models 2-3 change with $t$, and their time paths along with 95\% confidence intervals are displayed in Figure \ref{fig: fig1}. For both models, the $ACRW(t)$'s are inverted U-shaped with peaks around time period 5. Nevertheless, the time series variations are relatively small, as compared with their cross sectional variations, indicating that the average causal effects of aid on growth is consistently insignificant across time, provided the model specifications are true.
{\spacingset{1.4}
\begin{table}[htbp]
  \centering 
  \caption{Aggregate Causal Effects of Aid on Growth under FEC-STI and SCIA-II} 
  \label{tab_exgo} 
\begin{tabular}{lccc}
\hline\hline
& \multicolumn{3}{c}{\textit{Dependent variable: Growth}} \\ 
\cline{2-4} 
& (1) & (2) & (3)\\ 
\hline
\multirow{2}{*}{Aid} & 0.045 & $-$0.053 & 0.296 \\ 
& (0.057) & (0.115) & (0.378) \\ 
\multirow{2}{*}{Aid$^2$} &  & 0.003 & 0.002 \\ 
&  & (0.004) & (0.003) \\  
\multirow{2}{*}{Aid$\times$Log(initGDPPC)}  &  &  & $-$0.049 \\ 
&  &  & (0.063) \\ 
\multirow{2}{*}{Log(initGDPPC)} & $-$2.833$^{***}$ & $-$2.879$^{***}$ & $-$2.734$^{***}$ \\ 
  & (0.631) & (0.683) & (0.582) \\  
\multirow{2}{*}{Assassinations} & $-$0.017 & $-$0.014 & $-$0.002 \\ 
  & (0.171) & (0.167) & (0.171) \\ 
\multirow{2}{*}{Ethnic$\times$Assassinations} & $-$0.680 & $-$0.680 & $-$0.703 \\ 
  & (0.754) & (0.743) & (0.749) \\ 
\multirow{2}{*}{Lag of M2/GDP} & $-$0.012 & $-$0.012 & $-$0.013 \\ 
  & (0.011) & (0.012) & (0.011) \\ 
\multirow{2}{*}{$ACRW^{*}$} & 0.045 & $-$0.031 & $-$0.068 \\ 
& (0.057) & (0.116) & (0.144) \\ 
\hline
Observations & 801 & 801 & 801 \\ 
R$^{2}$ & 0.076 & 0.078 & 0.079 \\ 
Adjusted R$^{2}$ & $-$0.068 & $-$0.067 & $-$0.067 \\ 
F Statistic & \makecell{11.349$^{***}$ \\(df = 5; 692)} & \makecell{9.759$^{***}$\\ (df = 6; 691)} & \makecell{8.505$^{***}$ \\(df = 7; 690)} \\ 
\hline \hline
\end{tabular}
\begin{tablenotes}
\footnotesize
\item Standard errors are in parentheses and are clustered at the country level except for $ACRW^{*}$ at columns (2)-(3), where the standard errors are calculated by bootstrap with 1000 replications. $^{*}$p$<$0.1; $^{**}$p$<$0.05; $^{***}$p$<$0.01\\
\end{tablenotes}
\end{table} 
}

Now we assume a FEC-SEI model along with the SCIA-I assumption. The functional forms of the alternative econometrics models for the observed data are the same as above. The moment conditions are now different. Specifically, for any $t=1,\dots, T$, $\varepsilon_{it}$'s are conditionally mean independent of the right-hand side variables at all periods $s\leq t$, and the coefficients in Models 1-3 are estimated by GMM (with predetermined variables). Similarly as above, after the coefficients are estimated, we can obtain estimates of $ACRW_{t}, t=1,...,T$ and $ACRW^{*}$ from the estimators proposed in Section 3, and obtain their standard errors by nonparametric panel bootstrap. 

\begin{figure}
\centering
\includegraphics[scale=0.9]{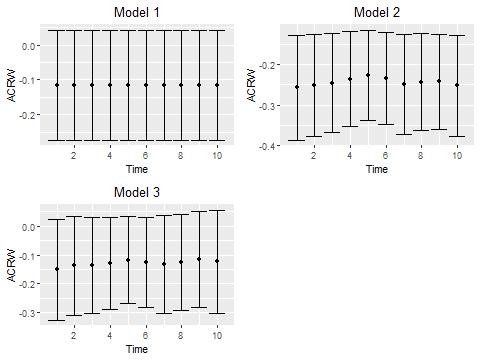}
\caption{Time paths of ACRW for Models 1-3 under FEC-SEI and SCIA-I}
\label{fig: fig2}
\end{figure}

The estimation results are displayed in Table 3. Under Model 1, the treatment effect is just $\beta_{1}$, and its estimate from data is -0.116 with a standard error of 0.080. The estimates of $ACRW^{*}$ for Models 2-3 are -0.245 and -0.130, respectively, and their bootstrap standard errors are 0.061 and 0.085, respectively. The $ACRW(t), t=1, \dots, T$, along with their 95\% confidence intervals, are displayed in Figure \ref{fig: fig2}. The situation is similar to that in Figure \ref{fig: fig1}, hence we omit the discussion of results. 
\begin{table}[htbp]
  \centering 
  \caption{Aggregate Causal Effects of Aid on Growth under FEC-SEI and SCIA-I} 
  \label{tab_predt} 
\begin{tabular}{lccc}
  \hline\hline 
  & \multicolumn{3}{c}{\textit{Dependent variable: Growth}} \\
  \cline{2-4} 
  & (1) & (2) & (3)\\ 
  \hline
  \multirow{2}{*}{Aid} & -0.116 & -0.282 & -0.633** \\ 
   & (0.080) & (0.181) & (0.239) \\ 
  \multirow{2}{*}{Aid$^2$} &  & 0.005 & 0.006† \\ 
   &  & (0.005) & (0.003) \\ 
  \multirow{2}{*}{Aid$\times$Log(initGDPPC)} &  &  & 0.061 \\ 
   &  &  & (0.047) \\
  \multirow{2}{*}{Log(initGDPPC)} & -4.396† & -4.317* & -4.975** \\ 
   & (2.346) & (1.785) & (1.608) \\ 
  \multirow{2}{*}{Assassinations} & -0.008 & -0.053 & -0.041 \\ 
   & (0.138) & (0.136) & (0.158) \\ 
  \multirow{2}{*}{Ethnic$\times$Assassinations} & -0.411 & -0.394 & -0.348 \\ 
   & (0.613) & (0.652) & (0.640) \\ 
  \multirow{2}{*}{Lag of M2/GDP} & -0.001 & 0.005 & 0.000 \\ 
   & (0.018) & (0.023) & (0.019) \\ 
  \multirow{2}{*}{$ACRW^{*}$} & -0.116 & -0.245 & -0.130 \\ 
  & (0.080) & (0.061) & (0.085) \\ 
\hline
Observations & 801 & 801 & 801 \\ 
Hansen J test & 196836.85*** & 224819.74*** & 248980.11*** \\ 
Wald test &  160304.02*** & 184056.84*** & 257263.60*** \\ 
\hline\hline 
\end{tabular}
\begin{tablenotes}
\footnotesize
\item[1] Standard errors are in parentheses and are clustered at the country level except for $ACRW^{*}$ at columns (2)-(3), where the standard errors are calculated by bootstrap with 1000 replications. $^{*}$p$<$0.1; $^{**}$p$<$0.05; $^{***}$p$<$0.01\\
\end{tablenotes}
\end{table} 

Note that in this paper we don't discuss the model specification issue, i.e., how to make choice between FEC-SEI and FEC-STI, and of the function form of $\tau()$. Our focus is on the definition and identification of causal effects after a set of model specifications is determined.

\section{Discussions}
TWFE with DiD design has been the predominant causal inference method for social scientists. What new insights does this paper bring into this tradition? It's hard to summarize in a few lines, as TWFE with DiD is now experiencing rapid methodological innovations. Nevertheless, we would like to point out the following aspects of our paper that are different from the existing DiD literature.

First, we consider continuous treatment under a non-staggered adoption design. As pointed out by \citet{de2023two}: ``The estimators proposed by \citet{de2024difference} and \citet{callaway2024difference} can accommodate continuous treatments, but only if the design is staggered. Many continuous treatments, such as precipitations or trade tariffs, do not follow a staggered design: they can change multiple times and can both go up or down.'' Our paper is a reply to their call into this direction.

Second, our model allows for flexible modeling of treatment heterogeneity. The $\tau(\overline{d}_{it}, Z_{it})$ function accommodates many heterogeneous models, such as the quadratic in current treatment specification, the additive and multiplicative in current and past treatment specifications, as special cases. It is easy to see that we can also include covariates into the  $\tau()$, say in the form of interactions with the treatments.

Third, we propose a different set of identifying assumptions, i.e., the SCIAs, than the parallel trend assumptions (PTAs) in the DiD literature. Both SCIAs and PTAs are assumptions on the conditional distribution of the potential outcomes. Since potential outcomes are not observable, neither SCIAs nor PTAs can be validated by empirical data directly. In observational studies, we have to rely something beyond current data to evaluate the plausibility of those assumptions. And contextual knowledge is the first choice of that something. In the appendix of this paper, we discuss how to validate the SCIA assumptions using the directed acyclic graphs (DAGs), as the SCIAs as DAGs are formal representation of contextual knowledge, which are just a collection of inter-relationships among all variables in the system.

In the DiD literature, the common practice to validate the PTAs is to compare the trends (of the treated and untreated) use pre-treatment data. But it's not an easy task to convince readers that the pre-treatment pattern will necessarily persist in the after treatment periods. The key hurdle of any causal inference is the unobservability of potential outcomes. Using pre-treatment historical data does not solve this problem. The researcher has to use rhetoric techniques to make his/her extrapolation (from pre-treatment to post treatment) persuasive \citep{cunningham2021causal}. But often times such extrapolation might backfire. A recent example is \citet{jaeger2019tweet}'s reexamining of \citet{kearney2015media}. \citet{kahn2020promise} conclude that the PTAs tests are neither necessary nor sufficient conditions for DiD to work. \citet{bilinski2018nothing} raise another issue with the traditional testing of PTAs: the powers of the tests are often too low.

This is not to say that DAGs are perfect tools to test the identifying assumptions. The legitimacy of the DAG approach hinges on the accuracy of the proposed DAGs. If the proposed DAG deviates from the truth, the method may fail as well. But the DAG approach forces the researcher to explicitly display his/her contextual knowledge and implicit assumptions, which can then be debated, and the whole field can focus on something tangible.

\section{Conclusion}
The burgeoning literature on DiD centers around rescuing TWFE, perhaps the most widely used method for causal inference with panel data, from its frailty in the presence of treatment effects heterogeneity, which originates from either the difference in treatment timing (e.g., staggered adoption) or the plurality of the treatment levels (e.g., multi-valued or continuous treatment).

This paper outlines a framework to evaluate treatment effects in panel data with a continuous treatment that is applicable to a wide range of situations, where treatment can change in either direction between any two consecutive periods.  Our specification of the treatment effects, i.e., the $\tau(\bar{d}_{it}, Z_{it})$ function, is flexible enough to allow not only for linearity in $d_{it}$, in which case treatment effects are homogeneous and TWFE works, but also for nonlinearity in $d_{it}$, additivity or  multiplicativeness in nonlinear functions of $d_{it}$ and past treatments such as $d_{i,t-1}$, in which case TWFE does not work and we have provided a new set of estimators for the aggregate average causal responses. The identifying assumptions, i.e., the conditions that translate models of potential outcomes to models of observable outcomes, are seemingly stronger than but essentially similar to those used in the DiD/TWFE literature, and are testable with a graphical tool, i.e., DAGs.

Throughout the paper our causal estimands are constructed using average causal responses (i.e., $ACR_{t}(\bar{d}_{it})$'s) as building blocks. We briefly discussed using other measures as building blocks such as the average treatment effects (i.e., $ATE_{t}(\bar{d}_{it},\bar{d}_{it}^{*})$). We have skipped other possible building blocks, such as the average treatment effects on the treated and their slopes. Though average causal responses are among ``the most natural building blocks'' to adopt in case of a continuous treatment, we have all too many other plausible candidates to use, and the researcher might need to decide which choice is best suited for his/her context \citep{callaway2024difference}, and when a different choice is needed, similar calculation and estimation can be conducted.

In this paper, we formulate the causal inference problem (i.e., the causal estimands and estimators, etc.) using the potential outcome framework and propose to evaluate the authenticity of the key identifying assumptions (i.e., the sequentially independence assumptions) with the causal graphs. This type of hybrid approach of combining the clarity of algebraic exposition and the abundance of statistical techniques of the former approach with the flexibility and intuitiveness of the latter approach has gained increasing popularity among applied researchers and become the fashion of recent textbooks in causal inference \citep[e.g.][etc.]{morgan2014counterfactuals,hernan2024causal,cunningham2021causal}. A reexmination of previous studies with this hybrid approach is a promising avenue for future research.

\clearpage
\bibliography{bibliography.bib}

\clearpage
\setcounter{equation}{0}
\setcounter{theorem}{0}
\setcounter{table}{0}
\setcounter{figure}{0}
\renewcommand{\theequation}{S.\arabic{equation}}
\renewcommand{\thetheorem}{S.\arabic{theorem}}
\renewcommand{\theassumption}{S.\arabic{assumption}}
\renewcommand{\thecorollary}{S.\arabic{corollary}}
\renewcommand{\thetable}{S.\arabic{table}}
\renewcommand{\thefigure}{S.\arabic{figure}}
\begin{appendices}
\section{Examples of treatment effect heterogeneity specifications}
The simplest functional form for $\tau(\overline{d}_{it}, Z_{it})$ is $\tau(\overline{d}_{it}, Z_{it})=\tau d_{it}$, where $\tau$ is an unknown parameter. This is the homogeneous treatment effects situation. It is easy to see that in this case
\[
ACR_{t}(\overline{d}_{it})=ACRW_{t}=ACRW^{*}=\tau,
\]
i.e., all levels of average causal responses are equal to the constant parameter $\tau$, and can be directly estimated from either (\ref{PDM}) by GMM or (\ref{STM}) by LSDV (i.e., TWFE), depending on which causal model and identifying assumptions we choose. Note also that in this case the standard errors of the estimators of $ACRW_{t}$ and $ACRW^{*}$ can be extracted from the estimation procedures directly.

A slightly more general form for $\tau(\overline{d}_{it}, Z_{it})$ is $\tau(\overline{d}_{it}, Z_{it})=\tau_{1} d_{it}+\tau_{2} d_{it}^{2}$, where $\tau_{1}, \tau_{2}$ are unknown parameters. Now we have heterogeneous treatment effects. Specifically,
\[
  ACR_{t}(\overline{d}_{it})=\tau_{1} +2\tau_{2} d_{it},
\]
and
\[
  \begin{aligned}
    ACRW_{t}&=\tau_{1} +2\tau_{2} \E[D_{it}], \\
    ACRW^{*}&=\frac{1}{T}\sum_{t=1}^{T}ACRW_{t}  =\tau_{1} +2\tau_{2} \sum_{t=1}^{T}\E[D_{it}]
  \end{aligned}
\]
Let $\hat{\tau}_{1}, \hat{\tau}_{2}$ be the estimators for $\tau_{1}, \tau_{2}$, respectively.\footnote{$\tau_{1}, \tau_{2}$ are estimated by GMM if we are using model (\ref{PDM}), and are estimated by LSDV (i.e., TWFE) if we are using model (\ref{STM}).} A consistent estimator for $ACRW_{t}$ is
\[
  \widehat{ACRW}_{t}=\hat{\tau}_{1} +2\hat{\tau}_{2}\hat{\E}[D_{it}]=\hat{\tau}_{1} +2\hat{\tau}_{2}\frac{1}{N}\sum_{i=1}^{N}D_{it},
\]
and a consistent estimator for $ACRW^{*}$ is
\[
  \widehat{ACRW}^{*}=\hat{\tau}_{1} +2\hat{\tau}_{2}\sum_{t=1}^{T}\hat{\E}[D_{it}]=\hat{\tau}_{1} +2\hat{\tau}_{2}\frac{1}{N}\sum_{t=1}^{T}\sum_{i=1}^{N}D_{it}.
\]

We can also introduce the interaction between $\overline{d}_{it}$ and $Z_{it}$ into the functional form for $\tau(\overline{d}_{it}, Z_{it})$, which is just the example considered in the main text.

Let's now discuss the situation where $  \tau(\overline{d}_{it}, Z_{it})$ depends on not only the current treatment $d_{it}$ but historical treatments $d_{1}, \dots, d_{t-1}$. For simplicity we assume that $\tau(\overline{d}_{it}, Z_{it})$ depends on $d_{it}$ and $d_{t-1}$, which means that ceteris paribus, potential outcomes are affected by the current period and the last period treatments, but not by the earlier period treatments,and there is no interaction effects between the treatments and the covariate $Z_{it}$.

If the treatment effects of different periods are additive, then we can assume an additive functional form for $\tau(\overline{d}_{it}, Z_{it})$, such as
\[
  \tau(\overline{d}_{it}, Z_{it})=\tau_{1}d_{it}+\tau_{2}d_{t-1},
\]
in which case
\[
  ACR_{t}(\overline{d}_{it})=\frac{\partial ATE_{t}(\overline{d}_{it})}{\partial d_{it}}=\tau_{1}
\]
hence
\[
  ACRW_{t}=ACRW^{*}=\tau_{1}.
\]
So under a linear additive model of treatment effects, $\tau_{1}$ represents the aggregate treatment effects, and we are back to the homogeneous treatment effects scenario discussed above.

A nonlinear additive model is
\[
  \tau(\overline{d}_{it}, Z_{it})=\tau_{1}f(d_{it})+\tau_{2}g(d_{t-1})
\]
where $f,g$ are nonlinear functions. Now
\[
  ACR_{t}(\overline{d}_{it})=\tau_{1}f'(d_{it})
\]
and
\[
  \begin{aligned}
  ACRW_{t}&=\E[\tau_{1}f'(D_{it})]=\tau_{1}\E[f'(D_{it})] \\
  ACRW^{*}&=\frac{1}{T-1}\sum_{t=2}^{T}ACRW_{t}=\frac{\tau_{1}}{T-1}\sum_{t=2}^{T}  \E[f'(D_{it})]
  \end{aligned}
\]
whose estimators are
\[
  \begin{aligned}
  \widehat{ACRW}_{t}&=\hat{\tau}_{1}\frac{1}{N}\sum_{i=1}^{N}f'(D_{it}) \\
  \widehat{ACRW}^{*}&=\hat{\tau}_{1}\cdot\frac{1}{N(T-1)}\sum_{i=1}^{N}\sum_{t=2}^{T}f'(D_{it})
  \end{aligned}
\]

If the treatment effects of different periods are multiplicative, then we can assume a multiplicative functional form for $\tau(\overline{d}_{it}, Z_{it})$, such as
\[
  \tau(\overline{d}_{it}, Z_{it})=\tau d_{it}d_{t-1},
\]
in which case
\[
  ACR_{t}(\overline{d}_{it})=\frac{\partial ATE_{t}(\overline{d}_{it})}{\partial d_{it}}=\tau d_{t-1}
\]
The estimation of aggregate treatment effects can be similarly derived.

We can also consider non-linearly multiplicative models, such as
\[
  \tau(\overline{d}_{it}, Z_{it})=\tau_{1}f(d_{it})g(d_{t-1})+\tau_{2}F(d_{it})G(d_{t-1})
\]
where $f,G,F,G$ are polynomials. A simple situation is
\[
  \tau(\overline{d}_{it}, Z_{it})=\tau_{1}d_{it}d_{t-1}+\tau_{2}d_{it}^{2}d_{t-1}
\]
Now
\[
  ACR_{t}(\overline{d}_{it})=\frac{\partial ATE_{t}(\overline{d}_{it})}{\partial d_{it}}=\tau_{1}d_{t-1}+2\tau_{2}d_{it}d_{t-1}
\]
Hence
\[
  ACRW_{t}=\E[D_{t-1}(\tau_{1}+2\tau_{2}D_{it})]
\]
and
\[
  \begin{aligned}
    ACRW^{*}&=\frac{1}{T-1}\sum_{t=2}^{T}ACRW_{t}=\frac{\tau_{1}}{T-1}\sum_{t=2}^{T}  \E[D_{t-1}(\tau_{1}+2\tau_{2}D_{it})]\\
    &=\tau_{1}\frac{1}{T-1}\sum_{t=2}^{T} \E[D_{t-1}] + 2\tau_{2}\frac{1}{T-1}\sum_{t=2}^{T} \E[D_{t-1}D_{it}]
  \end{aligned}
\]
Their estimators are
\[
  \begin{aligned}
    \widehat{ACRW}_{t}&=\hat{\tau}_{1}\hat{\E}[D_{t-1}] + 2\hat{\tau}_{2} \hat{\E}[D_{t-1}D_{it}]\\
    &=\hat{\tau}_{1}\frac{1}{N}\sum_{i=1}^{N}D_{i,t-1}+2\hat{\tau}_{2}\frac{1}{N}\sum_{i=1}^{N}D_{i,t-1}D_{it} \\
    \widehat{ACRW}^{*}&=\hat{\tau}_{1}\frac{1}{T-1}\sum_{t=2}^{T} \hat{\E}[D_{t-1}] + 2\hat{\tau}_{2}\frac{1}{T-1}\sum_{t=2}^{T} \hat{\E}[D_{t-1}D_{it}]\\
    &=\hat{\tau}_{1}\frac{1}{N(T-1)}\sum_{i=1}^{N}\sum_{t=2}^{T}D_{i,t-1}+2\hat{\tau}_{2}\frac{1}{N(T-1)}\sum_{i=1}^{N}\sum_{t=2}^{T} D_{i,t-1}D_{it}
  \end{aligned}
\]
respectively. 

\section{Proofs}
\subsection{Proof of Theorem \ref{th1}}
By the definition of $\mathbf{R}_i$ and $\mathbf{R}_i^*$, we have $\mathbf{R}_i^* = \mathbf{R}_i - \frac{1}{N}\sum_{i=1}^N \mathbf{R}_i$. Note that 
\[
\begin{aligned}
    \frac{1}{N}\sum_{i=1}^N \mathbf{W}_i^\prime \mathbf{R}_i^* & = \frac{1}{N}\sum_{i=1}^N \mathbf{W}_i^\prime \Big(\mathbf{R}_i - \frac{1}{N}\sum_{i=1}^N \mathbf{R}_i\Big) = \frac{1}{N}\sum_{i=1}^N \mathbf{W}_i^\prime \mathbf{R}_i - \frac{1}{N}\sum_{i=1}^N \mathbf{W}_i^\prime \frac{1}{N}\sum_{i=1}^N \mathbf{R}_i \\
    & \pto \E(\mathbf{W}_i^\prime\mathbf{R}_i) - \E(\mathbf{W}_i^\prime)\E(\mathbf{R}_i) = \mathbf{G}.
\end{aligned}
\]
Thus, it follows from the continuous mapping theorem that 
\begin{equation}
\label{pt}
\bigg[\bigg(\frac{1}{N}\sum_{i=1}^N \mathbf{W}_i^\prime \mathbf{R}_i^*\bigg)^\prime \boldsymbol{\Omega} \bigg(\frac{1}{N}\sum_{i=1}^N \mathbf{W}_i^\prime \mathbf{R}_i^*\bigg) \bigg]^{-1} \pto (\mathbf{G}^\prime\boldsymbol{\Omega}\mathbf{G})^{-1}.
\end{equation}
Then, denoting $\Delta \boldsymbol{\varepsilon}_i = (\Delta \varepsilon_{i2},\ldots,\Delta \varepsilon_{iT})^\prime$ and $\Delta \boldsymbol{\varepsilon}_i^* = (\Delta \varepsilon_{i2}^*,\ldots,\Delta \varepsilon_{iT}^*)^\prime$, we also have 
\[
\begin{aligned}
    \frac{1}{N}\sum_{i=1}^N \mathbf{W}_i^\prime \Delta \boldsymbol{\varepsilon}_i^* & = \frac{1}{N}\sum_{i=1}^N \mathbf{W}_i^\prime \Big(\Delta \boldsymbol{\varepsilon}_i - \frac{1}{N}\sum_{i=1}^N \Delta \boldsymbol{\varepsilon}_i \Big) \\
    & = \frac{1}{N}\sum_{i=1}^N \mathbf{W}_i^\prime \Delta \boldsymbol{\varepsilon}_i - \frac{1}{N}\sum_{i=1}^N \mathbf{W}_i^\prime \frac{1}{N}\sum_{i=1}^N \Delta \boldsymbol{\varepsilon}_i.
\end{aligned}
\]
By the multivariate CLT,
\[
\sqrt{N}\left(\begin{bmatrix}
    \frac{1}{N} \sum_{i=1}^N \mathbf{W}_i^\prime \Delta \boldsymbol{\varepsilon}_i\\
    \frac{1}{N} \sum_{i=1}^N \mathrm{Vec}(\mathbf{W}_i^\prime) \\
    \frac{1}{N} \sum_{i=1}^N \Delta \boldsymbol{\varepsilon}_i\\
    \frac{1}{N} \sum_{i=1}^N \mathbf{Q}_i
\end{bmatrix} - \begin{bmatrix}
    \mathbf{0} \\
    \\E[\mathrm{Vec}(\mathbf{W}_i^\prime)] \\
    \mathbf{0}\\
    \E(\mathbf{Q}_i)
\end{bmatrix}\right) \dto N\left(\mathbf{0},\mathrm{Var}\begin{pmatrix}
    \mathbf{W}_i^\prime \Delta \boldsymbol{\varepsilon}_i\\
    \mathrm{Vec}(\mathbf{W}_i^\prime)\\
    \Delta \boldsymbol{\varepsilon}_i\\
    \mathbf{Q}_i
\end{pmatrix}\right) = N(\mathbf{0},\boldsymbol{\Sigma}).
\]
By the definition of $h(\cdot)$, we have  
\[
h\left(\begin{bmatrix}
    \frac{1}{N} \sum_{i=1}^N \mathbf{W}_i^\prime \Delta \boldsymbol{\varepsilon}_i\\
    \frac{1}{N} \sum_{i=1}^N \mathrm{Vec}(\mathbf{W}_i^\prime) \\
    \frac{1}{N} \sum_{i=1}^N \Delta \boldsymbol{\varepsilon}_i\\
    \frac{1}{N} \sum_{i=1}^N \mathbf{Q}_i 
\end{bmatrix}\right) = \begin{bmatrix}
    \frac{1}{N}\sum_{i=1}^N \mathbf{W}_i^\prime \Delta \boldsymbol{\varepsilon}_i - \frac{1}{N}\sum_{i=1}^N \mathbf{W}_i^\prime \frac{1}{N}\sum_{i=1}^N \Delta \boldsymbol{\varepsilon}_i \\
    \frac{1}{N} \sum_{i=1}^N \mathbf{Q}_i
\end{bmatrix}.
\]
After observing that 
\[
h\left(\begin{bmatrix}
    \mathbf{0} \\
    \\E[\mathrm{Vec}(\mathbf{W}_i^\prime)] \\
    \mathbf{0}\\
    \E(\mathbf{Q}_i)
\end{bmatrix}\right) = \begin{bmatrix}
    \mathbf{0} \\
    \E(\mathbf{Q}_i)
\end{bmatrix},
\]
it follows from the multivariate delta method that 
\[
\sqrt{N}\left(\begin{bmatrix}
    \frac{1}{N}\sum_{i=1}^N \mathbf{W}_i^\prime \Delta \boldsymbol{\varepsilon}_i^*\\
    \frac{1}{N}\sum_{i=1}^N \mathbf{Q}_i
\end{bmatrix} - \begin{bmatrix}
    \mathbf{0} \\
    \E(\mathbf{Q}_i)
\end{bmatrix} \right) \dto N(\mathbf{0},\mathbf{B}\boldsymbol{\Sigma}\mathbf{B}^\prime).
\]
Then, it follows from \eqref{pt} and the Slutsky's theorem that 
\begin{align*}
    & \sqrt{N}\left(\begin{bmatrix}
        \widehat{\phi}^G\\
        \frac{1}{N} \sum_{i=1}^N \mathbf{Q}_i
    \end{bmatrix}-\begin{bmatrix}
        \phi \\
        \E(\mathbf{Q}_i)
    \end{bmatrix}\right) \\
    & = \begin{bmatrix}
        \bigg[\bigg(\frac{1}{N}\sum_{i=1}^N \mathbf{W}_i^\prime \mathbf{R}_i^*\bigg)^\prime \boldsymbol{\Omega} \bigg(\frac{1}{N}\sum_{i=1}^N \mathbf{W}_i^\prime \mathbf{R}_i^*\bigg) \bigg]^{-1} \bigg(\frac{1}{N}\sum_{i=1}^N \mathbf{W}_i^\prime \mathbf{R}_i^*\bigg)^\prime \boldsymbol{\Omega} & \mathbf{0} \\
        \mathbf{0} & \mathbf{I}
    \end{bmatrix} \\
    & \hspace{10cm}\begin{bmatrix}
    \frac{1}{\sqrt{N}}\sum_{i=1}^N \mathbf{W}_i^\prime \Delta \boldsymbol{\varepsilon}_i^*\\
    \frac{1}{\sqrt{N}}\sum_{i=1}^N [\mathbf{Q}_i - \E(\mathbf{Q}_i)]
\end{bmatrix}\\
    & \dto \begin{bmatrix}
        (\mathbf{G}^\prime\boldsymbol{\Omega}\mathbf{G})^{-1}\mathbf{G}^\prime\boldsymbol{\Omega} & \mathbf{0} \\
        \mathbf{0} & \mathbf{I}
    \end{bmatrix}N(\mathbf{0},\mathbf{B}\boldsymbol{\Sigma}\mathbf{B}^\prime),
\end{align*}
that is, 
\[
\sqrt{N}\left(\begin{bmatrix}
    \widehat{\phi}^G\\
    \frac{1}{N} \sum_{i=1}^N \mathbf{Q}_i
\end{bmatrix}-\begin{bmatrix}
    \phi \\
    \E(\mathbf{Q}_i)
\end{bmatrix}\right) \dto N(\mathbf{0},\mathbf{H}\mathbf{B}\boldsymbol{\Sigma}\mathbf{B}^\prime\mathbf{H}^\prime).
\]
Finally, it follows from the multivariate delta method again that 
\[
\begin{aligned}
    & \sqrt{N}\bigg(f\Big(\widehat{\phi}^G,\frac{1}{N}\sum_{i=1}^N\mathbf{Q}_i\Big)-f\Big(\phi,\E(\mathbf{Q}_i)\Big)\bigg) = \sqrt{N}(\widehat{CE}-CE) \\
    & \dto N\Big(0,\nabla_f(\phi,\E(\mathbf{Q}_i))^\prime\mathbf{H}\mathbf{B}\boldsymbol{\Sigma}\mathbf{B}^\prime\mathbf{H}^\prime\nabla_f(\phi,\E(\mathbf{Q}_i))\Big).
\end{aligned}
\]
Q.E.D.

\subsection{Proof of Theorem \ref{th2}}
By the definition of $\ddot{\mathbf{D}}_i$ and $\dot{\mathbf{D}}_i$, we have $\ddot{\mathbf{D}}_i = \dot{\mathbf{D}}_i - \frac{1}{N}\sum_{i=1}^N \dot{\mathbf{D}}_i$. Note that 
\begin{align*}
  \frac{1}{N} \sum_{i=1}^N \ddot{\mathbf{D}}_i^\prime \ddot{\mathbf{D}}_i & = \frac{1}{N} \sum_{i=1}^N \bigg(\dot{\mathbf{D}}_i - \frac{1}{N}\sum_{i=1}^N \dot{\mathbf{D}}_i\bigg)^\prime \bigg(\dot{\mathbf{D}}_i - \frac{1}{N}\sum_{i=1}^N \dot{\mathbf{D}}_i\bigg) \\
  & = \frac{1}{N} \sum_{i=1}^N \dot{\mathbf{D}}_i^\prime \dot{\mathbf{D}}_i - \frac{1}{N}\sum_{i=1}^N \dot{\mathbf{D}}_i^\prime\frac{1}{N}\sum_{i=1}^N \dot{\mathbf{D}}_i \\
  & \pto \E(\dot{\mathbf{D}}_i^\prime \dot{\mathbf{D}}_i) - \E(\dot{\mathbf{D}}_i^\prime)\E(\dot{\mathbf{D}}_i) = \mathbf{U}.
\end{align*}
Therefore, it follows from the continuous mapping theorem that 
\begin{equation}
\label{pt2}
\bigg(\frac{1}{N} \sum_{i=1}^N \ddot{\mathbf{D}}_i^\prime \ddot{\mathbf{D}}_i\bigg)^{-1} \pto \mathbf{U}^{-1}.
\end{equation}
Moreover, denoting $\ddot{\boldsymbol{\varepsilon}}_i = (\ddot{\varepsilon}_{i1},\ldots,\ddot{\varepsilon}_{iT})^\prime$ and $\dot{\boldsymbol{\varepsilon}}_i=(\dot{\varepsilon}_{i1},\ldots,\dot{\varepsilon}_{iT})^\prime$, we have 
\begin{align*}
\frac{1}{N} \sum_{i=1}^N \ddot{\mathbf{D}}_i^\prime \ddot{\boldsymbol{\varepsilon}}_i & = \frac{1}{N} \sum_{i=1}^N\bigg(\dot{\mathbf{D}}_i - \frac{1}{N}\sum_{i=1}^N \dot{\mathbf{D}}_i\bigg)^\prime\bigg(\dot{\boldsymbol{\varepsilon}}_i-\frac{1}{N}\sum_{i=1}^N \dot{\boldsymbol{\varepsilon}}_i\bigg)\\
& = \frac{1}{N}\sum_{i=1}^N \dot{\mathbf{D}}_i^\prime \dot{\boldsymbol{\varepsilon}}_i - \frac{1}{N}\sum_{i=1}^N \dot{\mathbf{D}}_i^\prime\frac{1}{N}\sum_{i=1}^N \dot{\boldsymbol{\varepsilon}}_i.
\end{align*}
By the multivariate CLT, 
\[
\sqrt{N}\left(\begin{bmatrix}
    \frac{1}{N} \sum_{i=1}^N \dot{\mathbf{D}}_i^\prime \dot{\boldsymbol{\varepsilon}}_i\\
    \frac{1}{N} \sum_{i=1}^N \mathrm{Vec}(\dot{\mathbf{D}}_i^\prime) \\
    \frac{1}{N} \sum_{i=1}^N \dot{\boldsymbol{\varepsilon}}_i\\
    \frac{1}{N} \sum_{i=1}^N \mathbf{Q}_i
\end{bmatrix} - \begin{bmatrix}
    \mathbf{0} \\
    \\E[\mathrm{Vec}(\dot{\mathbf{D}}_i^\prime)] \\
    \mathbf{0}\\
    \E(\mathbf{Q}_i)
\end{bmatrix}\right) \dto N\left(\mathbf{0},\mathrm{Var}\begin{pmatrix}
    \dot{\mathbf{D}}_i^\prime \dot{\boldsymbol{\varepsilon}}_i\\
    \mathrm{Vec}(\dot{\mathbf{D}}_i^\prime)\\
    \dot{\boldsymbol{\varepsilon}}_i\\
    \mathbf{Q}_i
\end{pmatrix}\right) = N(\mathbf{0},\boldsymbol{\Xi}).
\]
By the definition of $g(\cdot)$, 
\[
g\left(\begin{bmatrix}
    \frac{1}{N} \sum_{i=1}^N \dot{\mathbf{D}}_i^\prime \dot{\boldsymbol{\varepsilon}}_i\\
    \frac{1}{N} \sum_{i=1}^N \mathrm{Vec}(\dot{\mathbf{D}}_i^\prime) \\
    \frac{1}{N} \sum_{i=1}^N \dot{\boldsymbol{\varepsilon}}_i\\
    \frac{1}{N} \sum_{i=1}^N \mathbf{Q}_i
\end{bmatrix}\right) = \begin{bmatrix}
    \frac{1}{N}\sum_{i=1}^N \dot{\mathbf{D}}_i^\prime \dot{\boldsymbol{\varepsilon}}_i - \frac{1}{N}\sum_{i=1}^N \dot{\mathbf{D}}_i^\prime\frac{1}{N}\sum_{i=1}^N \dot{\boldsymbol{\varepsilon}}_i \\
    \frac{1}{N} \sum_{i=1}^N \mathbf{Q}_i
\end{bmatrix}.
\]
Observing that 
\[
g\left(\begin{bmatrix}
    \mathbf{0} \\
    \\E[\mathrm{Vec}(\dot{\mathbf{D}}_i^\prime)] \\
    \mathbf{0}\\
    \E(\mathbf{Q}_i)
\end{bmatrix}\right) = \begin{bmatrix}
    \mathbf{0} \\
    \E(\mathbf{Q}_i)
\end{bmatrix},
\]
it follows from the multivariate delta method that 
\[
\sqrt{N}\left(\begin{bmatrix}
    \frac{1}{N}\sum_{i=1}^N \ddot{\mathbf{D}}_i^\prime \ddot{\boldsymbol{\varepsilon}}_i\\
    \frac{1}{N}\sum_{i=1}^N \mathbf{Q}_i
\end{bmatrix} - \begin{bmatrix}
    \mathbf{0} \\
    \E(\mathbf{Q}_i)
\end{bmatrix} \right) \dto N(\mathbf{0},\mathbf{C}\boldsymbol{\Xi}\mathbf{C}^\prime).
\]
Then, it follows from \eqref{pt2} and the Slutsky's theorem that 
\begin{align*}
    & \sqrt{N}\left(\begin{bmatrix}
        \widehat{\phi}^O\\
        \frac{1}{N} \sum_{i=1}^N \mathbf{Q}_i
    \end{bmatrix}-\begin{bmatrix}
        \phi \\
        \E(\mathbf{Q}_i)
    \end{bmatrix}\right) \\
    & = \begin{bmatrix}
        \bigg(\frac{1}{N} \sum_{i=1}^N \ddot{\mathbf{D}}_i^\prime \ddot{\mathbf{D}}_i\bigg)^{-1} & \mathbf{0} \\
        \mathbf{0} & \mathbf{I}
    \end{bmatrix} \begin{bmatrix}
    \frac{1}{\sqrt{N}}\sum_{i=1}^N \ddot{\mathbf{D}}_i^\prime \ddot{\boldsymbol{\varepsilon}}_i\\
    \frac{1}{\sqrt{N}}\sum_{i=1}^N [\mathbf{Q}_i - \E(\mathbf{Q}_i)]
\end{bmatrix}\\
    & \dto \begin{bmatrix}
        \mathbf{U}^{-1} & \mathbf{0} \\
        \mathbf{0} & \mathbf{I}
    \end{bmatrix}N(\mathbf{0},\mathbf{C}\boldsymbol{\Xi}\mathbf{C}^\prime) = N(\mathbf{0},\mathbf{J}\mathbf{C}\boldsymbol{\Xi}\mathbf{C}^\prime\mathbf{J}^\prime).
\end{align*}
Finally, it follows from the multivariate delta method again that 
\[
\begin{aligned}
    & \sqrt{N}\bigg(f\Big(\widehat{\phi}^O,\frac{1}{N}\sum_{i=1}^N\mathbf{Q}_i\Big)-f\Big(\phi,\E(\mathbf{Q}_i)\Big)\bigg) = \sqrt{N}(\widehat{CE}-CE) \\
    & \dto N\Big(0,\nabla_f(\phi,\E(\mathbf{Q}_i))^\prime\mathbf{J}\mathbf{C}\boldsymbol{\Xi}\mathbf{C}^\prime\mathbf{J}^\prime\nabla_f(\phi,\E(\mathbf{Q}_i))\Big).
\end{aligned}
\]
Q.E.D.

\subsection{Proof of Theorem \ref{ths1}}
Under the DPFEC-SEI or DPFEC-STI, we can only use GMM to estimate $\phi$ and build plug-in estimators for causal estimands. Therefore, the estimation procedure is identical to that under the model FEC-SEI except that the moment conditions are different. Consequently, after redefining the matrices as detailed in Appendix D, the proof of this theorem is identical to that of Theorem \ref{th1}, and is thus omitted for the sake of simplicity.

\section{Validating the identifying assumptions}
The SCIA-I and the SCIA-II assumptions are vital assumptions in the identification of causal effects. It's natural to ask how to verify whether these assumptions are valid in practice. If our data come from a sequentially randomized experiment, then it is easy to verify that both the SCIA assumptions hold. In this paper we are mainly concerned with observational studies where randomization is not present. Frankly speaking, validation of the SCIA is theoretically challenging, as it involves thinking about the unobservable potential outcomes. As \citet{pearl2018book} put it, ``Unfortunately, I have yet to find a single person who can explain what ignorability (i.e., the CIA) means in a language spoken by those who need to make this assumption or assess its plausibility in a given problem.'' \citet{hernan2024causal} comment on this issue in a similar manner: ``Answering this question is difficult because thinking in terms of conditional exchangeability (i.e., the CIA) is often not intuitive in complex causal systems.''

The theory of directed acyclic graph (DAG) provides an equivalent characterization of the CIA that is more friendly to reasoning and validation. Here we give a brief introduction to some of the basic concepts related to DAG. For detailed exposition of DAG related concepts and results, see \citet{pearl2016causal} and \citet{hernan2024causal}.

The DAG is a graphical representation of a structural causal model (SCM) where the relationships among variables within a finite system are delineated by a set of equations. Variables in the SCM are represented by nodes in a graph. For any pairs of variables $X$ and $Y$, there is an arrow (also called ``directed edge'') pointing from $X$ to $Y$ as long as there exists at least one structural equation for which $Y$ is the left-hand side variable and $X$ is one of the right-hand side variables. Put it differently, arrows (directed edges) represent causal effects not mediated by other variables in the system. The collection of all nodes and arrows among nodes is a \textit{directed graph}. A \textit{path} is a sequence of arrows that connect two nodes. A \textit{direct path} is a path in which all arrows point in the same direction. A node is a \textit{descendant} of another node if it can be reached by a directed path. A \textit{cycle} is a direct path that emanates from a node and also terminates at the same node. A directed graph without cycles is called a \textit{directed acyclic graph} (DAG).

In a given DAG, there are three possible configurations for a path with three nodes, say $X, Y$ and $Z$: a \textit{chain} ($X\rightarrow Z\rightarrow Y$), a \textit{fork} ($X\leftarrow Z\rightarrow Y$), and an \textit{inverted fork} ($X\rightarrow Z\leftarrow Y$, also known as a \textit{collider}). A path $p$ is \textit{blocked }by a set of nodes $\mathcal{Z}=\{Z\}$ if and only if (1) $p$ contains a chain ($X\rightarrow Z\rightarrow Y$) or a fork ($X\leftarrow Z\rightarrow Y$), with $Z\in \mathcal{Z}$; or (2) $p$ contains an inverted fork ($X\leftarrow Z\leftarrow Y$), and that neither $Z$ nor its descendants is contained in $\mathcal{Z}$.

Let $T$ and $Y$ be the treatment variable and the outcome variable respectively, in a DAG. A \textit{backdoor path} between $T$ and $Y$ is a path connecting $T$ and $Y$ that contains an arrow into $T$. A set of variables $X$ satisfies the \textit{backdoor criterion} relative to $(T,Y)$ if no node in $X$ is a descendant of $T$, and $X$ blocks every backdoor path between $T$ and $Y$.

According to \citet[][pp. 342-344]{pearl2009causality}, if we knows the true DAG that generated the data, then under certain regularity conditions, the CIA holds, i.e., $Y(t)\perp T~|~X, ~~ \forall t\in \mathcal{T}$, if and only if $X$ satisfies the backdoor criterion related to the treatment effect of $T$ on $Y$ (a similar proof is provided by Hernán and Robins, 2020, pp.85).\footnote{Note that unlike the potential outcome framework, the DAG framework does not use the concept of potential outcome. The two frameworks are logically equivalent, though (see \citealp[][Chapter 7]{pearl2009causality}, and \citealp[][pp. 125-126]{pearl2016causal}).} With this result, the assessment of whether the CIA holds is equivalent to the assessment of whether $X$ satisfies the backdoor criterion related to the treatment effect of $T$ on $Y$ in the DAG, assuming that the DAG is correctly specified. Since in a given DAG, the backdoor criterion can be checked directly, the problem now hinges on the reliability of the proposed DAG.

Now how do we validate the sequentially conditional independence assumptions such as the SCIA-I and the SCIA-II? The idea is simple: evaluate whether CIA holds at each time period by the DAG at that period.\footnote{\citet{hernan2024causal} Chapter 19-21 discussed how to verify SCIAs in longitudinal medical studies with DAGs. Note that in \citet{hernan2024causal} outcomes for each individual are only observed at the end of study, which is different from the panel data structure we are considering in this paper.}

Figure S.1 displays a simple DAG with two periods. It is easy to verify that if the DAG is as displayed in Figure S.1, then according to the back-door criterion, the SCIA-I assumption holds. If in Figure S.1, the arrow from $Y_1$ to $D_2$ is removed, then SCIA-II holds.
\begin{figure}[htbp]
	\centering
	\begin{tikzpicture}[>=Latex]
		\node (1) at (0,0) {$\mathbf{X}_{1}$};
		\node (2) [right = of 1] {$D_{1}$};
		\node (3) [right = of 2] {$Y_{1}$};
		\node (4) [right = of 3] {$\mathbf{X}_{2}$};
		\node (5) [right = of 4] {$D_{2}$};
		\node (6) [right = of 5] {$Y_{2}$};
		
		\draw [->] (1) to (2);
		\draw [->] (1) to[bend left=20] (3);
		\draw [->] (1) to[bend left=30] (4);
		\draw [->] (1) to[bend left=50] (5);	
		\draw [->] (1) to[bend left=60] (6);	
		\draw [->] (2) to (3);
		\draw [->] (2) to[bend right=30] (5);
    \draw [->] (2) to[bend right=40] (6);
		\draw [->] (3) to[bend left=40] (5);	
		\draw [->] (4) to (5);	
		\draw [->] (4) to[bend left=50] (6);		
		\draw [->] (5) to (6);		
	\end{tikzpicture} 
	\caption{A Simple DAG with Two Periods}
\end{figure}
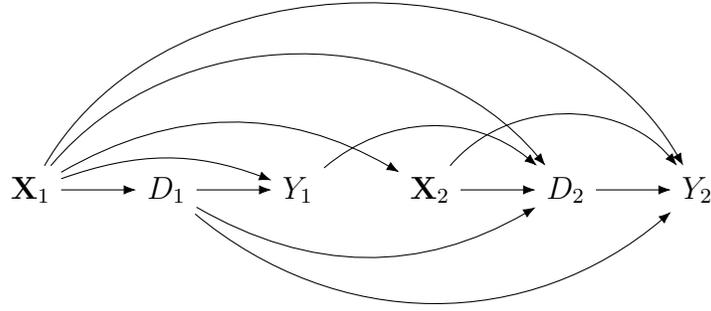

Now let's introduce unobservable individual and time fixed effects. First let's introduce individual effects $U$ into Figure S.1.  Let's first assume that $U$ does not affect the treatment variable $D$, which is reasonable if we are considering treatments that apply to group of individuals, but is probably untenable if we are considering personalized treatment. The DAG is displayed in Figure S.2.

\begin{figure}[ht]
	\centering
	\begin{tikzpicture}[>=Latex]
		\node (1) at (0,0) {$\mathbf{X}_{1}$};
		\node (2) [right = of 1] {$D_{1}$};
		\node (3) [right = of 2] {$Y_{1}$};
		\node (4) [right = of 3] {$\mathbf{X}_{2}$};
		\node (5) [right = of 4] {$D_{2}$};
		\node (6) [right = of 5] {$Y_{2}$};
		\node (7) [below = of 1] {$U$};
		
		\draw [->] (1) to  (2);
		\draw [->] (1) to[bend left=20]  (3);
		\draw [->] (1) to[bend left=30]  (4);
		\draw [->] (1) to[bend left=50]  (5);	
		\draw [->] (1) to[bend left=60]  (6);	
		\draw [->] (2) to (3);
		\draw [->] (2) to[bend right=30]  (5);
		\draw [->] (2) to[bend right=40]  (6);		
		\draw [->] (3) to[bend left=40]  (5);	
		\draw [->] (4) to  (5);	
		\draw [->] (4) to[bend left=50]  (6);		
		\draw [->] (5) to  (6);	
		
		\draw [->] (7) to (1);
		\draw [->] (7) to[bend right=30]  (3);
		\draw [->] (7) to[bend right=40]  (4);		
		\draw [->] (7) to[bend right=50]  (6);		
	\end{tikzpicture}
	\caption{A Two-period DAG with Individual Effects}
\end{figure}
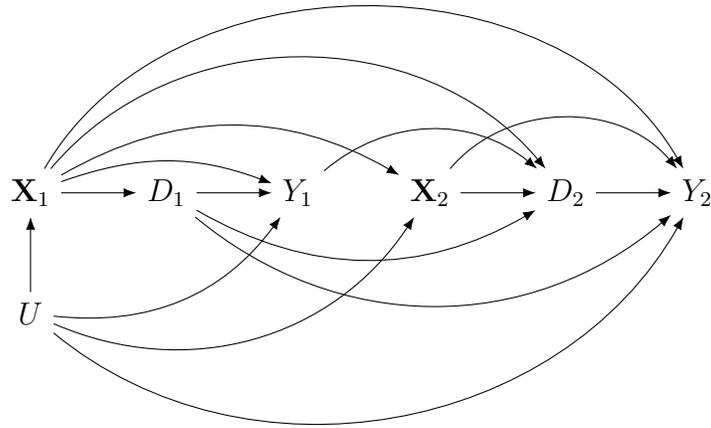
It is easy to verify that if the DAG is as displayed in Figure S.2, then according to the back-door criterion, SCIA-I holds. If the arrow from $Y_1$ to $D_2$ is removed in Figure S.2, then SCIA-II holds.

Now let's consider time fixed effects $V_{1}, V_{2}$. If we would like to assume that time fixed effects are just fixed parameters, like the econometrics tradition, then we don't need to add them into the DAGs, as they don't affect the causal relationship. If we assume that time fixed effects are unobservable  random variables, then similar as situation of the individual effects, SCIAs still hold as long as there is no arrow from $V_{1}, V_{2}$ to the treatment variables $D_{1}, D_{2}$ \citep[][\S19.4]{hernan2024causal}.

\begin{figure}[ht]
	\centering
	\begin{tikzpicture}[>=Latex]
		\node (1) at (0,0) {$\mathbf{X}_{1}$};
		\node (2) [right = of 1] {$D_{1}$};
		\node (3) [right = of 2] {$Y_{1}$};
		\node (4) [right = of 3] {$\mathbf{X}_{2}$};
		\node (5) [right = of 4] {$D_{2}$};
		\node (6) [right = of 5] {$Y_{2}$};
		\node (7) [below = of 1] {$U$};
		\node (8) [above = of 1] {$V_{1}$};
		\node (9) [below = of 4] {$V_{2}$};
		
		\draw [->] (1) to (2);
		\draw [->] (1) to[bend left=20]  (3);
		\draw [->] (1) to[bend left=30]  (4);
		\draw [->] (1) to[bend left=50]  (5);	
		\draw [->] (1) to[bend left=60]  (6);	
		\draw [->] (2) to (3);
		\draw [->] (2) to[bend right=23]  (5);
		\draw [->] (2) to[bend right=30]  (6);		
		\draw [->] (3) to[bend left=40]  (5);	
		\draw [->] (4) to  (5);	
		\draw [->] (4) to[bend left=50]  (6);		
		\draw [->] (5) to  (6);	
		
		\draw [->] (7) to (1);
		\draw [->] (7) to[bend right=30]  (3);
		\draw [->] (7) to[bend right=40]  (4);		
		\draw [->] (7) to[bend right=50]  (6);	
		\draw [->] (8) to (1);
		\draw [->] (8) to[bend left=30]  (3);
		\draw [->] (9) to (4);
		\draw [->] (9) to[bend right=30] (6);	    	    		
	\end{tikzpicture}
	\caption{A Two-period DAG with Two-way Effects}
\end{figure}
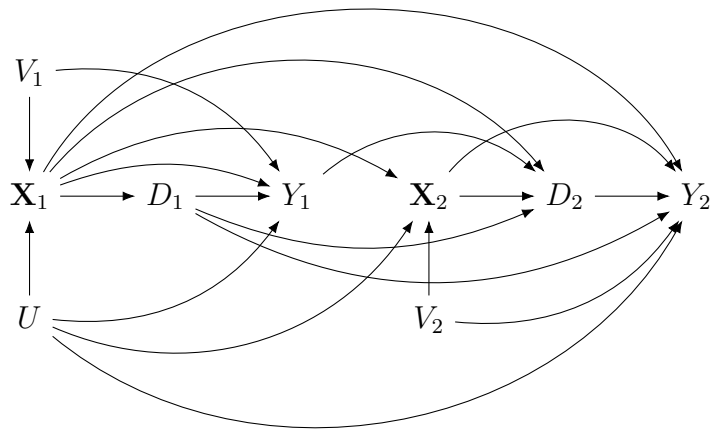

In the above we have discussed the situation that there are unobseravable confounders and we didn't include them in the equations. But since we actually have already included unobseravable confounders in our regression equations such as (\ref{PDM}) and (\ref{STM}), they are no longer unmeasured confounders, hence the SCIAs must hold if our DAG without $U,V$ can be represented as Figure S.1. The analysis of DAGs with more periods is similar hence omitted.

The introduction of the DAGs does not solve the causal identification problem in observational data as if by magic. In fact, it just pushes the causal investigation to a higher level: the structural causal model, which is built on contextual knowledge. Without additional information encapsulated in the contextual knowledge, causal inference in observational data is always on shaky ground. If the research community has reached consensus on the contextual knowledge, then we can build DAGs based on the agreed contextual knowledge. Yet, it is not rare that the research community does not have consensus on the contextual knowledge for a particular causal inference problem. Nevertheless, under general scenarios, researchers do have consensus on certain part of the contextual knowledge, which can be used as the basis to build data-consistent DAGs.  

\section{Results for dynamic panel data models}
In this appendix, we extended the results for static panel data models to dynamic ones. 

\textbf{Causal estimands.} When dynamic panel models for the potential outcomes are considered, the subpopulation whose average treatment effects we might be interested in can also be characterized by the lagged outcomes. Therefore, the causal estimand defined in \eqref{estm1} can be extended to 
\[
\begin{aligned}
ATE_{t}(\overline{d}_{it},\overline{d}_{it}^{*};G_t,G_t^*,S_{t},B_{t})=\E\Big[Y_{it}(\overline{d}_{it})-Y_{it}(\overline{d}_{it}^{*}) \mid Y_{i,t-1}(\overline{d}_{t-1}) \in G_t, Y_{i,t-1}(\overline{d}_{t-1}^*) \in G_t^*, \\
\overline{\mathbf{X}}_{it}\in S_{t}, \overline{D}_{i,k(t)}\in B_{t} \Big],  
\end{aligned}
\]
where $G_t$ is a subset of the support of $Y_{i,t-1}(\overline{d}_{t-1})$, and $G_t^*$ is a subset of the support of $Y_{i,t-1}(\overline{d}_{t-1}^*)$. Note that for simplicity, the dynamic panel models considered in this appendix only contain the first-order lagged outcome, so we do not condition on higher-order lagged outcome in $ATE_{t}(\overline{d}_{it},\overline{d}_{it}^{*};G_t,G_t^*,S_{t},B_{t})$. However, all the results in this appendix are compatible with dynamic panel models and average treatment effects that include higher-order lagged outcomes. Moreover, as in the main text, $G_t$, $G_t^*$, $S_t$, $B_t$ can all be set to the support of $Y_{i,t-1}(\overline{d}_{t-1})$, $Y_{i,t-1}(\overline{d}_{t-1})$, $\overline{\mathbf{X}}_{it}$, $\overline{D}_{i,k(t)}$, respectively, so $ATE_{t}(\overline{d}_{it},\overline{d}_{it}^{*};G_t,G_t^*,S_{t},B_{t})$ reduces to the $ATE_{t}(\overline{d}_{it},\overline{d}_{it}^{*})$ again. 

Regarding the average causal response $ACR_t(\overline{d}_{it})$ and aggregated average treatment effects $ATEW_t$, $ATEW^*$, $ACRW_t$, $ACRW^*$ defined in the main text, they remain appropriate causal estimands whose identification and estimation can be similarly achieved under the dynamic panel scenario.

\textbf{Potential outcome models.} Next, we extend the models FEC-SEI and FEC-STI investigated in the main text to their dynamic panel versions. We first propose the dynamic panel fixed effects causal model with sequentially mean independent errors (DPFEC-SEI),
\[
\E\Big[Y_{it}(\overline{d}_{it}) \mid Y_{i,t-1}(\overline{d}_{t-1}), \overline{\mathbf{X}}_{it}, U_{i}, \overline{V}_{t}\Big]=\gamma Y_{i,t-1}(\overline{d}_{t-1}) + \beta^\prime\mathbf{X}_{it}+\tau(\overline{d}_{it}, Z_{it})+U_{i}+V_{t}, 
\]
which can be reformulated as 
\[
\begin{gathered}
Y_{it}(\overline{d}_{it})=\gamma Y_{i,t-1}(\overline{d}_{t-1}) + \beta^\prime\mathbf{X}_{it}+\tau(\overline{d}_{it}, Z_{it})+U_{i}+V_{t}+\varepsilon_{it}(\overline{d}_{it}), \\
\\E[\varepsilon_{it}(\overline{d}_{it}) \mid Y_{i,t-1}(\overline{d}_{t-1}), \overline{\mathbf{X}}_{it}, U_{i}, \overline{V}_{t}] = 0.
\end{gathered}
\]

Then we propose the dynamic panel fixed effects causal model with strictly mean independent errors (DPFEC-STI),
\[
\E\Big[Y_{it}(\overline{d}_{it}) \mid Y_{i,t-1}(\overline{d}_{t-1}), \overline{\mathbf{X}}_{iT}, U_{i}, \overline{V}_{T}\Big]=\gamma Y_{i,t-1}(\overline{d}_{t-1}) + \beta^\prime\mathbf{X}_{it}+\tau(\overline{d}_{it}, Z_{it})+U_{i}+V_{t}, 
\]
which can be equivalently rewritten as
\[
\begin{gathered}
Y_{it}(\overline{d}_{it})=\gamma Y_{i,t-1}(\overline{d}_{t-1}) + \beta^\prime\mathbf{X}_{it}+\tau(\overline{d}_{it}, Z_{it})+U_{i}+V_{t}+\varepsilon_{it}(\overline{d}_{it}), \\
\\E[\varepsilon_{it}(\overline{d}_{it}) \mid Y_{i,t-1}(\overline{d}_{t-1}), \overline{\mathbf{X}}_{iT}, U_{i}, \overline{V}_{T}] = 0.
\end{gathered}
\]
The only difference between these two models and the models in the main text is the incorporation of the term $\gamma Y_{i,t-1}(\overline{d}_{i,t-1})$, which captures the effect of the first-order lagged outcome on the current outcome. The interpretations of other components within these models remain the same as those in the main text. 

\textbf{Identifying assumptions.} To identify these models, identifying assumptions proposed in the main text (i.e., SCIA-I and SCIA-II) should be extended to their dynamic panel forms as well. Specifically, to identify DPFEC-SEI, the following dynamic panel sequentially conditional independence assumption (DPSCIA-I) is required, 
\begin{equation}
\label{dpcia1}
\{\mathbf{X}_{ij}, V_{j}, Y_{ij}(\overline{d}_{ij})\}_{j=t+1}^{T},Y_{it}(\overline{d}_{it}) \perp D_{it} \mid \overline{Y}_{i,t-1}(\overline{d}_{i,t-1}), \overline{\mathbf{X}}_{it}, \overline{D}_{i,t-1}=\overline{d}_{i,t-1}, \overline{V}_{t}, U_i, ~ \forall t=1,\dots, T.
\end{equation}
Under DPSCIA-I, it can be proven that 
\[
\E\Big[Y_{it}(\overline{d}_{it}) \mid \overline{Y}_{i,t-1}(\overline{d}_{i,t-1}), \overline{\mathbf{X}}_{it}, \overline{V}_{t}, U_i\Big] = \E\Big[Y_{it} \mid \overline{Y}_{i,t-1}, \overline{\mathbf{X}}_{it}, \overline{D}_{it}=\overline{d}_{it}, \overline{V}_{t}, U_i\Big].
\]
Therefore, if DPFEC-SEI also holds, we have 
\[
\E\Big[Y_{it} \mid \overline{Y}_{i,t-1},\overline{\mathbf{X}}_{it}, \overline{D}_{it}=\overline{d}_{it}, \overline{V}_{t}, U_{i}\Big] = \gamma Y_{i,t-1}(\overline{d}_{t-1}) + \beta^\prime \mathbf{X}_{it}+\tau(\overline{d}_{it}, Z_{it})+U_{i}+V_{t},
\]
or equivalently,
\[
\E\Big[Y_{it} \mid \overline{Y}_{i,t-1},\overline{\mathbf{X}}_{it}, \overline{D}_{it}, \overline{V}_{t}, U_{i}\Big] = \gamma Y_{i,t-1} + \beta^\prime \mathbf{X}_{it}+\tau(\overline{D}_{it}, Z_{it})+U_{i}+V_{t},
\]
which yields the following model for observed outcomes,
\begin{equation}
\label{dp1}
\begin{gathered}
Y_{it}= \gamma Y_{i,t-1} + \beta'\mathbf{X}_{it}+\tau(\overline{D}_{it}, Z_{it})+U_{i}+V_{t}+ \varepsilon_{it},\\
\\E[\varepsilon_{it} \mid \overline{Y}_{i,t-1}, \overline{\mathbf{X}}_{it}, \overline{D}_{it}, \overline{V}_{t}, U_{i}] = 0.
\end{gathered}
\end{equation}
The parameters in this model are typically estimated by the GMM, and we will show the estimation procedure later.

Next, to identify DPFEC-STI, we modify \eqref{SCIA-II} to the following condition,
\[
Y_{it}(\overline{d}_{it}) \perp \overline{D}_{i\{j>t\}} \mid Y_{i,t-1}(\overline{d}_{i,t-1}), \overline{\mathbf{X}}_{iT}, \overline{D}_{it}=\overline{d}_{it}, \overline{V}_{T}, U_i,  \quad \forall t=1,\dots, T-1,
\]
which, together with \eqref{dpcia1}, will be referred to as DPSCIA-II.

Under DPSCIA-II, it can be proven that 
\[
\E\Big[Y_{it}(\overline{d}_{it}) \mid \overline{Y}_{i,t-1}(\overline{d}_{i,t-1}), \overline{\mathbf{X}}_{iT}, \overline{V}_{T}, U_i\Big] = \E\Big[Y_{it} \mid \overline{Y}_{i,t-1}, \overline{\mathbf{X}}_{iT}, \overline{D}_{iT}=\overline{d}_{iT}, \overline{V}_{T}, U_i\Big].
\]
Consequently, if DPFEC-STI further holds, we have the following model for observed outcomes, 
\begin{equation}
\label{dp2}
\begin{gathered}
Y_{it}= \gamma Y_{i,t-1} + \beta'\mathbf{X}_{it}+\tau(\overline{D}_{it}, Z_{it})+U_{i}+V_{t}+ \varepsilon_{it},\\
\\E[\varepsilon_{it} \mid \overline{Y}_{i,t-1}, \overline{\mathbf{X}}_{iT}, \overline{D}_{iT}, \overline{V}_{T}, U_{i}] = 0,
\end{gathered}
\end{equation}
where the parameters will be estimated by the GMM again.

\textbf{Estimation.} Same as the scenario of static panel models, without loss of generality, suppose that $\tau(\overline{d}_{it}, Z_{it})$ has the following form 
\[
\tau(\overline{D}_{it},Z_{it}) = \tau_1 M_{it1} + \tau_2 M_{it2} + \cdots + \tau_S M_{itS}.
\]
Therefore, the parameters to be estimated are $\phi=(\gamma,\beta^\prime,\tau_1,\ldots,\tau_S)^\prime$.

To estimate $\phi$ under the model DPFEC-SEI, we perform differencing on \eqref{dp1} twice to eliminate the individual and time fixed effects,
\[
\Delta Y_{it}^* = \gamma \Delta Y_{i,t-1}^* + \beta^\prime \Delta \mathbf{X}_{it}^* + \tau_1 \Delta M_{it1}^* + \tau_2 \Delta M_{it2}^* + \cdots + \tau_S \Delta M_{itS}^* + \Delta \varepsilon_{it}^*,
\]
where $\Delta Y_{it}^* = \Delta Y_{it} - \frac{1}{N}\sum_{i=1}^N \Delta Y_{it}$, and other variables are defined similarly. Based on the properties of errors in \eqref{dp1}, it is easy to derive that 
\[
\\E[\Delta \varepsilon_{it}^* \mid \overline{Y}_{i,t-1},\overline{\mathbf{X}}_{i,t-1}, \overline{D}_{i,t-1}] = 0,
\]
so we have the following moment conditions: for $i=1,\ldots,N$ and $t=3,\ldots,T$,
\[
\begin{aligned}
    \E\Big[\Big(Y_{i1},\ldots,Y_{i,t-2},&\mathbf{X}_{i1}^\prime,\ldots,\mathbf{X}_{i,t-1}^\prime, D_{i1},\ldots,D_{i,t-1}\Big)^\prime\\
    & \Big(\Delta Y_{it}^* - \gamma \Delta Y_{i,t-1}^* - \beta^\prime \Delta \mathbf{X}_{it}^* - \tau_1 \Delta M_{it1}^* - \cdots - \tau_S \Delta M_{itS}^*\Big)\Big] = \mathbf{0}.
\end{aligned}
\]
Then, define
\[
\mathbf{W}_i = \begin{bmatrix}
    (Y_{i1},\overline{\mathbf{X}}_{i2}^\prime,\overline{D}_{i2}^\prime) & \mathbf{0}^\prime & \cdots & \mathbf{0}^\prime \\
    \mathbf{0}^\prime & (\overline{Y}_{i2},\overline{\mathbf{X}}_{i3}^\prime,\overline{D}_{i3}^\prime) & \cdots & \mathbf{0}^\prime \\
    \vdots & \vdots & \ddots & \vdots \\
    \mathbf{0}^\prime & \mathbf{0}^\prime & \cdots & (\overline{Y}_{i,T-2},\overline{\mathbf{X}}_{i,T-1}^\prime,\overline{D}_{i,T-1}^\prime)
\end{bmatrix},
\]
and 
\[
\mathbf{R}_i^* = \begin{bmatrix}
    \Delta Y_{i,2}^* & \Delta \mathbf{X}_{i3}^{*\prime} & \Delta M_{i31}^* & \cdots & \Delta M_{i3S}^* \\
    \vdots & \vdots & \vdots & \ddots & \vdots \\
    \Delta Y_{i,T-1}^* & \Delta \mathbf{X}_{iT}^{*\prime} & \Delta M_{iT1}^* & \cdots & \Delta M_{iTS}^*
\end{bmatrix}.
\]
The GMM estimator $\widehat{\phi}^G$ for $\phi$ is obtained by  
\begin{equation}
\label{eqphi}
\widehat{\phi}^G = \bigg[\bigg(\frac{1}{N}\sum_{i=1}^N \mathbf{W}_i^\prime \mathbf{R}_i^*\bigg)^\prime \boldsymbol{\Omega} \bigg(\frac{1}{N}\sum_{i=1}^N \mathbf{W}_i^\prime \mathbf{R}_i^*\bigg) \bigg]^{-1} \bigg[\bigg(\frac{1}{N}\sum_{i=1}^N \mathbf{W}_i^\prime \mathbf{R}_i^*\bigg)^\prime \boldsymbol{\Omega} \bigg(\frac{1}{N}\sum_{i=1}^N \mathbf{W}_i^\prime \Delta \mathbf{Y}_i^* \bigg)\bigg],
\end{equation}
where $\boldsymbol{\Omega}$ is a positive definite weighting matrix. 

Next, we estimate $\phi$ under the model DPFEC-STI. Note that the model expression remains unchanged, but based on the properties of errors in \eqref{dp2}, the moment conditions become: for $i=1,\ldots,N$ and $t=3,\ldots,T$,
\[
\begin{aligned}
    \E\Big[\Big(Y_{i1},\ldots,Y_{i,t-2},&\mathbf{X}_{i1}^\prime,\ldots,\mathbf{X}_{i,T-1}^\prime, D_{i1},\ldots,D_{i,T-1}\Big)^\prime\\
    & \Big(\Delta Y_{it}^* - \gamma Y_{i,t-1} - \beta^\prime \Delta \mathbf{X}_{it}^* - \tau_1 \Delta M_{it1}^* - \cdots - \tau_S \Delta M_{itS}^*\Big)\Big] = \mathbf{0}.
\end{aligned}
\]
Defining  
\[
\mathbf{W}_i = \begin{bmatrix}
    (Y_{i1},\overline{\mathbf{X}}_{i,T-1}^\prime,\overline{D}_{i,T-1}^\prime) & \mathbf{0}^\prime & \cdots & \mathbf{0}^\prime \\
    \mathbf{0}^\prime & (\overline{Y}_{i2},\overline{\mathbf{X}}_{i,T-1}^\prime,\overline{D}_{i,T-1}^\prime) & \cdots & \mathbf{0}^\prime \\
    \vdots & \vdots & \ddots & \vdots \\
    \mathbf{0}^\prime & \mathbf{0}^\prime & \cdots & (\overline{Y}_{i,T-2},\overline{\mathbf{X}}_{i,T-1}^\prime,\overline{D}_{i,T-1}^\prime)
\end{bmatrix}
\]
and 
\[
\mathbf{R}_i^* = \begin{bmatrix}
    \Delta Y_{i,2}^* & \Delta \mathbf{X}_{i3}^{*\prime} & \Delta M_{i31}^* & \cdots & \Delta M_{i3S}^* \\
    \vdots & \vdots & \vdots & \ddots & \vdots \\
    \Delta Y_{i,T-1}^* & \Delta \mathbf{X}_{iT}^{*\prime} & \Delta M_{iT1}^* & \cdots & \Delta M_{iTS}^*
\end{bmatrix},
\]
the GMM estimator $\widehat{\phi}^G$ for $\phi$ is obtained by \eqref{eqphi} as well.

Finally, similar to the construction of estimators demonstrated in Section 3, by replacing $\phi$ and expectations in causal estimands with $\widehat{\phi}^G$ and corresponding sample averages, we obtain estimators of causal estimands under these two models. 

\textbf{Asymptotic theory.} Since we use GMM again (remember that we use GMM to estimate $\phi$ in FEC-SEI) to estimate $\phi$ in the dynamic panel models, the asymptotic theory is essentially the same as that stated in Theorem \ref{th1}. Specifically, the asymptotic distributions of the causal estimators are:
\begin{theorem}
\label{ths1}
Under DPFEC-SEI with DPSCIA-I, DPFEC-STI with DPSCIA-II, and under standard regularity conditions for GMM estimators, we have 
\[
\sqrt{N} (\widehat{CE} - CE) \dto N\Big(0,\nabla_f(\phi,\E(\mathbf{Q}_i))^\prime\mathbf{H}\mathbf{B}\boldsymbol{\Sigma}\mathbf{B}^\prime\mathbf{H}^\prime\nabla_f(\phi,\E(\mathbf{Q}_i))\Big),
\]
where we assume that $\mathbf{B}\boldsymbol{\Sigma}\mathbf{B}^\prime$ is positive definite and $\nabla_f(\phi,\E(\mathbf{Q}_i))^\prime\mathbf{H}\mathbf{B}\boldsymbol{\Sigma}\mathbf{B}^\prime\mathbf{H}^\prime\nabla_f(\phi,\E(\mathbf{Q}_i)) \allowbreak > 0$. The definition of the vectors and matrix is the same as before, except that $\mathbf{W}_i$ and $\mathbf{R}_i^*$ have been redefined for both DPFEC-SEI and DPFEC-STI.
\end{theorem}
\end{appendices}
\end{document}